\documentclass[12pt,preprint]{aastex}

\usepackage{lscape}
\usepackage{comment}

\shorttitle{A Likely Detection of Two Cold Planets in a Low-mag Event}
\shortauthors{Suzuki et al.}

\newcommand\ltsima{$\; \buildrel <\over\sim \;$}
\newcommand\simlt{\lower.5ex\hbox{\ltsima}}
\newcommand\gtsima{$\; \buildrel >\over\sim \;$}
\newcommand\simgt{\lower.5ex\hbox{\gtsima}}

\begin{document}


\title{A Likely Detection of a Two-Planet System in a Low Magnification Microlensing Event}


\author{D. Suzuki\altaffilmark{1,A}, D. P. Bennett\altaffilmark{2,3,A}, A. Udalski\altaffilmark{4,B}, I. A. Bond\altaffilmark{5,A}, T. Sumi\altaffilmark{6,A}, 
C. Han\altaffilmark{7}\\
 and \\
F. Abe\altaffilmark{8}, Y. Asakura\altaffilmark{8}, R.K. Barry\altaffilmark{9}, A. Bhattacharya\altaffilmark{2,3}, M. Donachie\altaffilmark{10}, M. Freeman\altaffilmark{10}, A. Fukui\altaffilmark{11}, Y. Hirao\altaffilmark{6,2}, Y. Itow\altaffilmark{8}, N. Koshimoto\altaffilmark{6},  M.C.A. Li\altaffilmark{10}, C. H. Ling\altaffilmark{5}, K. Masuda\altaffilmark{8}, Y. Matsubara\altaffilmark{8}, Y. Muraki\altaffilmark{8}, M. Nagakane\altaffilmark{6}, K. Onishi\altaffilmark{12}, H. Oyokawa\altaffilmark{8}, C. Ranc\altaffilmark{2}, N. J. Rattenbury\altaffilmark{10}, To. Saito\altaffilmark{13}, A. Sharan\altaffilmark{10}, D. J. Sullivan\altaffilmark{14}, P. J. Tristram\altaffilmark{15}, A. Yonehara\altaffilmark{16} \\
(the MOA Collaboration)\\
and \\
R. Poleski\altaffilmark{4,17}, P. Mr{\'o}z\altaffilmark{4}, J. Skowron\altaffilmark{4}, M.K. Szyma{\'n}ski\altaffilmark{4}, 
I. Soszy{\'n}ski\altaffilmark{4}, S. Koz{\l}owski\altaffilmark{4}, P. Pietrukowicz\altaffilmark{4},
{\L}. Wyrzykowski\altaffilmark{4}, K. Ulaczyk\altaffilmark{4} \\ (the OGLE Collaboration)
}


\altaffiltext{1}{Institute of Space and Astronautical Science, Japan Aerospace Exploration Agency, 3-1-1 Yoshinodai, Chuo, Sagamihara, Kanagawa 252-5210, Japan}
\altaffiltext{2}{Laboratory for Exoplanets and Stellar Astrophysics, NASA / Goddard Space Flight Center, Greenbelt, MD 20771, USA}
\altaffiltext{3}{Department of Astronomy, University of Maryland, College Park, MD 20742, USA}
\altaffiltext{4}{Warsaw University Observatory, Al. Ujazdowskie 4, 00-478 Warszawa, Poland}
\altaffiltext{5}{Institute of Information and Mathematical Sciences, Massey University, Private Bag 102-904, North Shore Mail Centre, Auckland, New Zealand}
\altaffiltext{6}{Department of Earth and Space Science, Graduate School of Science, Osaka University, 1-1 Machikaneyama, Toyonaka, Osaka 560-0043, Japan}
\altaffiltext{7}{Department of Physics, Chungbuk National University, Cheongju 28644, Korea}
\altaffiltext{8}{Institute for Space-Earth Environmental Research, Nagoya University, Furo-cho, Chikusa, Nagoya, Aichi 464-8601, Japan}
\altaffiltext{9}{Astrophysics Science Division, NASA/Goddard Space Flight Center, Greenbelt, MD 20771, USA}
\altaffiltext{10}{Department of Physics, University of Auckland, Private Bag 92019, Auckland, New Zealand}
\altaffiltext{11}{Okayama Astrophysical Observatory, National Astronomical Observatory, 3037-5 Honjo, Kamogata, Asakuchi, Okayama 719-0232, Japan}
\altaffiltext{12}{Nagano National College of Technology, Nagano 381-8550, Japan}
\altaffiltext{13}{Tokyo Metropolitan College of Industrial Technology, Tokyo 116-8523, Japan}
\altaffiltext{14}{School of Chemical and Physical Sciences, Victoria University, Wellington, New Zealand}
\altaffiltext{15}{University of Canterbury Mt. John Observatory, P.O. Box 56, Lake Tekapo 8770, New Zealand}
\altaffiltext{16}{Department of Physics, Faculty of Science, Kyoto Sangyo University, Kyoto 603-8555, Japan}
\altaffiltext{17}{Department of Astronomy, Ohio State University, 140 West 18th Avenue, Columbus, OH 43210, USA}
\altaffiltext{A}{Microlensing Observations in Astrophysics (MOA) Collaboration}
\altaffiltext{B}{Optical Gravitational Lensing Experiment (OGLE) Collaboration}


\begin{abstract}
We report on the analysis of a microlensing event OGLE-2014-BLG-1722 
that showed two distinct short term anomalies.
The best fit model to the observed light curves shows that 
the two anomalies are explained with two planetary mass ratio companions to the primary lens.
Although a binary source model is also able to explain the second anomaly,
it is marginally ruled out by 3.1 $\sigma$.
The 2-planet model indicates that the first anomaly was caused by planet ``b" 
with a mass ratio of $q = (4.5_{-0.6}^{+0.7}) \times 10^{-4}$ and projected 
separation in unit of the Einstein radius, $s = 0.753 \pm 0.004$. 
The second anomaly reveals planet ``c" with a mass ratio of 
$q_{2} = (7.0_{-1.7}^{+2.3}) \times 10^{-4}$ 
with $\Delta \chi^{2} \sim 170$ compared to the single planet model.
Its separation has two degenerated solutions: 
the separation of planet c is $s_{2} = 0.84 \pm 0.03$ and $s_{2} = 1.37 \pm 0.04$ for the close and wide models, respectively.
Unfortunately, this event dose not show clear finite source and microlensing parallax effects, 
thus we estimated the physical parameters of the lens system from Bayesian analysis.
This gives that 
the masses of planet b and c are 
$m_{\rm b} = 56_{-33}^{+51}\,M_{\oplus}$ and $m_{\rm c} = 85_{-51}^{+86}\,M_{\oplus}$, respectively, 
and they orbit a late type star with a mass of $M_{\rm host} = 0.40_{-0.24}^{+0.36}\,M_{\odot}$ 
located at $D_{\rm L} = 6.4_{-1.8}^{+1.3}\,\rm kpc$ from us.
The projected distance between the host and planets are $r_{\perp, \rm b} = 1.5 \pm 0.6\, \rm AU$ for planet b, 
and $r_{\perp, \rm c} = 1.7_{-0.6}^{+0.7}\, \rm AU$ and $r_{\perp, \rm c} = 2.7_{-1.0}^{+1.1}\, \rm AU$ for close and wide models of planet c.
If the 2-planet model is true, then 
this is the third multiple planet system detected by using the microlensing method, 
and the first multiple planet system detected in the low magnification events, 
which are dominant in the microlensing survey data.
The occurrence rate of multiple cold gas giant systems is estimated 
using the two such detections and a simple extrapolation of 
the survey sensitivity of 6 year MOA microlensing survey \citep{suz16} 
combined with the 4 year $\mu$FUN detection efficiency \citep{gou10}.
It is estimated that $6 \pm 2\,\%$ of stars host two cold giant planets.
\end{abstract}





\section{Introduction}
\label{sec_int}

Multiple planet systems are not uncommon in the planetary systems found by {\it Kepler} \citep{lis11,fab14}.
$22.4\,\%$ of discovered planetary systems by transit method are multiple planet systems\footnote{As of February 23, 2017 from http://exoplanet.eu/}. 
Transiting multiple systems provide much more information than transiting single systems, i.e., mutual inclination, orbital period ratio, and average density with the mass measurement by transit timing variation.
Such a rich information allows us to constrain the planet formation models.
Tightly packed coplanar multiple systems, such as Kepler-11 \citep{kep11} and 
TRAPPIST-1 \citep{gil17}, are 
thought to form in outer region than the current orbit and migrated inward through a disk.
Transiting single planet systems could be dynamically hotter and less orderly compared to the transiting multiple planet systems \citep{mor14}.
These studies have rapidly developed our understandings of architecture of close-in, hot multiple planet systems, 
but there is a huge parameter space waiting for our probes: cold, wide orbit multiple planet systems.

Radial Velocity method (RV) 
is sensitive to planets with wider orbits than transit, 
and it has detected a lot of multiple planet systems. 
The ratio of the RV multiple planet systems to the number of planetary systems in the RV sample is $22.2\,\%^{1}$, which is interestingly almost same as the transit sample.
However, the inclination of each planet is not able to be derived with the RV alone.
Also, finding wide orbit planets with the mass of below Saturn mass is difficult with current facilities. 
Direct imaging (DI) can find multiple planet systems if the system has young, bright, massive objects in very wide (mostly $\geq 10\,\rm AU$) orbit \citep{mar08}, but, again, detecting planets with sub-Jupiter mass or below is very difficult with the current instruments. 
The ratio of number of multiple systems to the number of whole systems detected by DI is $4.2\,\%^{1}$, which is probably reflecting the intrinsic low occurrence rate of multiple super Jupiters in wide orbit.

Microlensing does have a sensitivity down to Earth mass planets \citep{ben96} orbiting
beyond the snow line, at which ices including water ice condense to increase the surface density of solid materials by a factor of five in the protoplanetary disks \citep{ida05,kk08}.
Massive planets with very wide orbits \citep{ben12,pol14}, even for unbound ones \citep{sum11,mro17ffp}, are detectable by microlensing.
Thus, in principle, microlensing is the best tool to study the cold multiple planet systems, especially for small planets, 
which are not able to be probed by the other methods (including the astrometry by the GAIA mission).
The host star to planet mass ratio and normilized projected separation can be always directly measured from observed microlensing light curves. 
The planet mass and physical projected separation are also measurable for about half of planetary systems with combinations of measurements of the finite source effect, Earth orbital parallax effect \citep{mur11}, satellite parallax \citep{uda15_sp,str16} and high angular resolution images for the lens flux measurement \citep{fuk15,kos17} and resolving the lens and source \citep{ben15hst,bat15}.
However, measuring the orbital parameters could be difficult for most of the planetary microlensing events.
Nevertheless, the inclination and eccentricity of the first multiple planets found by microlensing, OGLE-2006-BLG-109Lbc, were constrained with the assumption of coplanar orbit \citep{ben10}.

So far, only two multiple planet systems have been found by microlensing against 49 planetary systems, 
so the observed multiple ratio is $4.1\,\%^{1}$, which is almost same as the DI sample, and much smaller than the transit and RV results.
There are two reasons for the low observed multiple ratio in the microlensing sample.
First, the rate of very high-magnification events, which is the major 
microlensing channel to detect multiple planet systems, is low, although 
the detection efficiency through this channel is high \citep{gau98_multi}.
Second, the detection efficiency for multiple Neptune mass (or below) planet system is not high, 
and the occurrence rate of such systems would be also not very high (We do not yet know about the occurrence rate for multiple Neptune mass planets).
The two multiple planet systems were found through the high-magnification channel.
OGLE-2006-BLG-109L is a system of Jupiter and Saturn analogues 
with orbits approximately half of Jupiter and Saturn orbits, respectively,
around the host with a half of the solar mass \citep{gau08, ben10}.
The second multiple planet system is OGLE-2012-BLG-0026L \citep{han13}, which is a sub-Jupiter and a sub-Saturn system with a solar mass host \citep{jp16}.
For the both systems, the masses were measured with the robust parallax effect measurement and further constraints from the high angular resolution images.
By contrast, low magnification events are expected to have a detection efficiency of $\mathcal{O}(1\%)$ or less\footnote{Actual average detection efficiency from the Microlensing Observations in Astrophysics 
\citep[MOA:][]{bon01, sum03}
survey in 2007--2012 \citep{suz16} is $\sim 0.01\,\%$ for two planet system with each mass ratio of $10^{-4}$, assuming that the detection efficiency of multiple planet system can be approximated by the product of detection efficiency to each planet \citep{gau98_multi,hanpark02}.} to detect multiple (two planet) system \citep{gau12}. 

In this paper, we report the analysis of OGLE-2014-BLG-1722, a low-magnification microlensing event 
showing two planetary anomalies which were caused by a lens system having two Saturn-mass planets.
The event time scale of this event is $\sim 24\,\rm days$, which is a typical value of microlensing events observed toward the Galactic bulge.
The mass of the lens system, $M$, and 
distance to the lens system, $D_{\rm L}$, are related to the event time scale by the relation,
\begin{equation}
t_{\rm E} = \frac{\theta_{\rm E}}{\mu_{\rm rel}}, ~~~~~ \theta_{\rm E } = \sqrt{{4GM\over c^2} \left(\frac{1}{D_{\rm L}} - \frac{1}{D_{\rm S}}\right)},
\end{equation}
where $\mu_{\rm rel}$ is the relative lens-source proper motion and $D_{\rm S}$ is the distance to the source star.
For most planetary events, we can measure the finite source 
effect (or source crossing time), leading to the measurement of the angular Einstein radius $\theta_{\rm E}$.
Furthermore, if we measure the microlensing parallax $\pi_{\rm E} \equiv \pi_{\rm rel}/\theta_{\rm E}$, $\pi_{\rm rel} = {\rm AU}(D_{\rm L}^{-1}-D_{\rm S}^{-1})$, we can robustly measure the mass of and distance to the lens system, assuming the source distance ($\sim 8\,\rm kpc$).
Unfortunately, 
neither of these effects was
not measured in this planetary microlensing event, thus we estimated the lens property with the Bayesian approach.
We describe observation and data reduction in Section \ref{sec_obs}.
In Section \ref{sec_model}, we show our light curve modeling.
In Section \ref{sec_phy}, we estimate the physical parameters:
the distance to the system, mass of the host star and two planets, and projected separation between the host and planets 
based on the Bayesian analysis.
We briefly discuss the validity of the second anomaly signal in Section \ref{sec_real}.
In Section \ref{sec_occ}, we estimate the occurrence rate of multiple cold gas giant planet systems.
In Section \ref{sec_sum}, we summarize the results and conclude.

\section{Observations and Data Reduction}
\label{sec_obs}





%

\begin{figure}
\begin{center}
\epsscale{.80}
\plotone{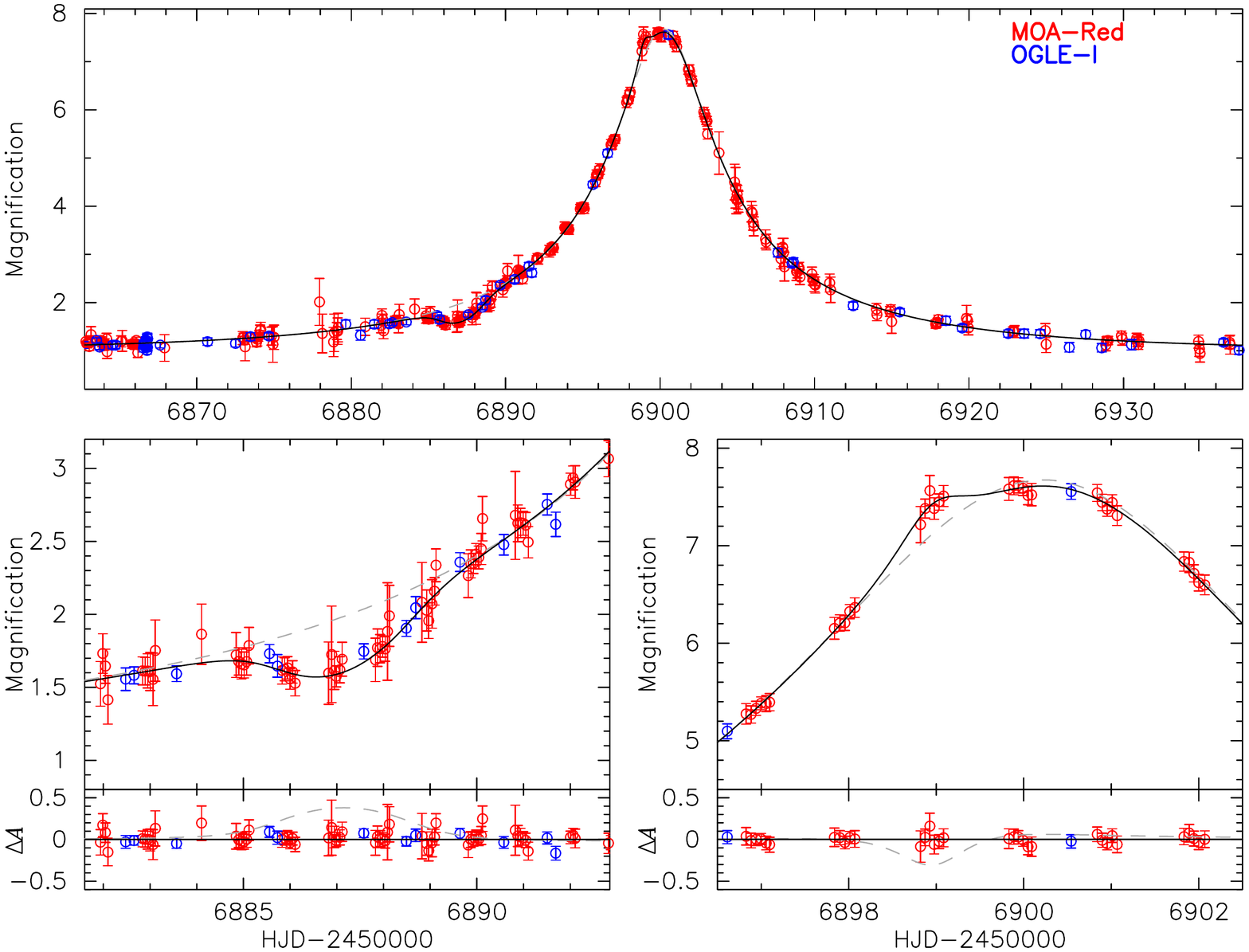}
\caption{Light curve of OGLE-2014-BLG-1722/MOA-2014-BLG-490 and the best fit 2-planet model. The left and right bottom panels show zoom-in of the anomalies caused by the planet b and c, respectively. 
Gray dashed line shows the best single lens fit.
MOA data points are binned into 1.2 hours, but unbinned data points are used for the light curve modeling.}
\label{fig:bestfit}
\end{center}
\end{figure}

The microlensing event OGLE-2014-BLG-1722 was alerted by the 
Optical Gravitational Lens Experiment 
\citep[OGLE:][]{uda03} collaboration 
using their real-time event detection system, Early Warning System (EWS) at $\rm HJD' \equiv HJD-2450000 = 6889.1$\footnote{14:18(UT), August 19, 2014} as a new microlensing event at $\rm (RA, Dec)_{2000} = (\rm 17^{h}55^{m}00^{s}\hspace{-3pt}.57, -31^{\circ}28'08''\hspace{-4pt}.6)$.
The event occurred in the OGLE-IV field \citep{uda15ogle4} BLG507 which is observed 1-3 times per night using the 1.4 $\rm deg^{2}$ camera on its 1.3 m Warsaw Telescope at the Las Campanas Observatory in Chile.
This microlensing event is also in the footprint of the MOA (Bond et al. 2001; Sumi et al.2003) survey, which uses the MOA-II 1.8 m telescope with a 2.2 $\rm deg^{2}$ field-of-view CCD camera 
MOA-cam3 \citep{sak08} at Mt. John University Observatory in New Zealand.
Following the OGLE alert, the MOA team discovered this event 
independently at $\rm HJD'=6890.0$ and labeled it as MOA-2014-BLG-490.
Since this event is found in one of the highest-cadence MOA fields, 
the images were taken more than 20 times per night.
The observed light curves of OGLE and MOA are plotted with the 
blue and red circles in Figure \ref{fig:bestfit}, where MOA data points are binned into 1.2 hours.

No one noticed in real-time the first anomaly occurred a few days before the discovery alerts, as well as the second, weak anomaly at $\rm HJD'=6899$.
The first anomaly was found about 20 days after the magnification peak.
Even at the peak, the event was not bright enough for observations using small telescopes.
Also, when the anomaly was found, it was clear that the anomaly had ended and the entire light curve was well covered by the OGLE and MOA teams.
Thus, no observations were conducted by the follow-up teams, and the OGLE and MOA survey observations were conducted with the normal cadence throughout the event.
As a result, this event can be classified as a planetary event discovered by pure surveys \citep{yee12,shv14,suz14,kos14,rat15,mro17}, which is a main stream of the current microlensing planet detection 
opposed to the survey+follow-up discoveries in the early time of microelnsing observation.

The OGLE and MOA data were at first reduced by using their difference image analysis (DIA) photometry pipelines \citep{uda03,bon01}, respectively.
After OGLE-2014-BLG-1722 was found to be a planetary event, 
the OGLE and MOA data were carefully re-reduced to precisely 
characterize the anomalies. 
The optimized centroid of the event position was used for the re-reduced OGLE light curve.
For the MOA images, we use a numerical kernel \citep{bra08} with a modification to allow for a spatial variation of the kernel across the sub-images centered on the event.
Also, we correct seeing, airmass and differential refraction correlations in the photometry, which can often be seen for the MOA images.

There could be possible systematics in the long baseline in OGLE and MOA data, which can sometimes affect the microlensing parallax measurements and other parameters.
To avoid this systematics, we use the photometry data in only 2014 and 2015 for the light curve modeling.
Since we want to estimate the accurate uncertainties for each fitting parameter, 
we normalize the error bars by using $\sigma'_{i} = k\sqrt{\sigma^{2}_{i} + e^{2}_{\rm min}}$, 
where $\sigma_{i}$ is the original error bar of the $i$th data point in magnitudes, and $k$ and $e_{\rm min}$ are the renormalizing parameters \citep{yee12}.
We expect that the cumulative of $\chi^{2}$ in each data point sorted with the magnification of the best fit model is a straight line if the data are normal distribution.
Thus, we choose $e_{\rm min}$ to make this cumulative $\chi^{2}$ distribution a straight line, and $k$ is chosen so that each data set gives $\chi^{2}/\rm dof = 1$.
We use $k = 1.305371$ and $e_{\rm min} = 0.00$ for OGLE data, and $k=1.139291$ and $e_{\rm min} = 0.00$ for MOA data, respectively, for the final result.

To double check the second anomaly feature, a different point spread function (PSF) is also used for the MOA data reduction. 
For this purpose, we reduced the MOA images by using DIA where we construct the PSF with DoPHOT package \citep{dophot,ben12}. 
The generated light curves with the DoPHOT PSF are usually used to measure the source star color, 
with which the angular Einstein radius $\theta_{\rm E}$ can be estimated if we measure the source crossing time $t_{\ast}$. 
In this event, however, the source crossing time was not able to be measured as the source trajectory did not cross the caustics or approach the cusps.
So, we use the DoPHOT PSF light curves only for light curve modeling as written in Section \ref{sec_real}.



\section{Light Curve Modeling}
\label{sec_model}

\subsection{1-Planet Model}
\label{sec_1pmodel}

The light curve of OGLE-2014-BLG-1722 is almost symmetric 
except for the short-term feature at $\rm HJD' \sim 6887$, 
a dip lasting for a few days (See Figure \ref{fig:bestfit}).
Considering the duration of the dip, the feature at a wing 
of the light curve is expected to be caused by a small mass ratio 
companion to the lens located inside of the Einstein radius.
Nevertheless, we investigate the vast parameter space to find the best fit model for the observed light curve.


A microlens light curve produced by a single mass is described by 
three parameters: the time of the peak magnification, $t_{0}$, 
the event time scale, $t_{\rm E}$, which is the time required for the 
Einstein radius crossing time, and the impact parameter divided by the angular Einstein radius, $u_{0}$.
If the lens object has a companion, three additional parameters are required: the mass ratio of the planet to the host star, $q\,(\equiv m_{\rm b}/M_{\rm host})$, the projected separation divided by the angular Einstein radius, $s$, and the angle between the source trajectory and planet-star axis, $\alpha$.
Most planetary events are characterized by the strong features of caustic crossings, at which the magnification of a point source diverges.
But, the actual source star has a finite angular size, so the magnification is limited such that a sharp magnification change is smoothed out.
Thus, modeling this rounded feature in a light curve allows us to measure the relative source star radius with an additional parameter, the angular source star radius relative to the angular Einstein radius, $\rho$, or the source radius crossing time, $t_{\ast} \equiv \rho t_{\rm E}$.

In addition to the above seven parameters, two linear parameters for 
each photometry data are required to describe the light curve: the flux 
from the source star, $F_{\rm S}$ and flux from the blended stars, $F_{\rm B}$.
Thus, the light curve model is given by $F(t) = A(t)F_{\rm S} + F_{\rm B}$, 
where $F(t)$ is the flux at a given time $t$ and $A(t) = A(t; t_{0}, t_{\rm E}, u_{0}, q, s, \alpha, \rho)$ is the magnification.

We use a variation of the Markov Chain Monte Carlo (MCMC) algorithm \citep{ver03} 
to search for the best fit model efficiently \citep{bennett-himag} with the re-reduced photometries.
We start to explore the parameter space with a grid for the parameters $q, s, \alpha$ that are sensitive to anomaly signals. 
We search for the other parameters by a downhill approach. 
The range of the grid for the mass ratio, $q$, is $10^{-4}$ -- $1$, 
which well covers the planetary and stellar binary mass ratio regime. 
As for the separation, we investigate $-0.5 < {\rm log}\,s < +0.5$ because microlensing is sensitive to the companions in this region. We use 22, 11, 40 grids for ${\rm log} s, {\rm log} q$ and $\alpha$, respectively.
After the grid search, we selected best 100 models as initial models to run the fitting code by letting each parameter free.
 
The best fit 1-planet model we found\footnote{This model is consistent with the first circulated planetary model by CH on September 18, 2014.} has a mass ratio of $q\sim 4\times 10^{-4}$, which corresponds to a sub-Saturn planet if we assume that the mass of the host star is $0.5 M_{\odot}$.
The separation is $s\sim 0.76$, which agrees with what we expect from the dip feature at the light curve wing.
The $\chi^{2}$ improvement for this planetary model over the single lens model is $\Delta \chi^{2}=1029$, so the planetary signal was significantly detected by the two survey teams. 
But, unfortunately, the relative source star radius, $\rho$, cannot be measured 
because the source did not cross the caustics.
The Einstein radius crossing time of $t_{\rm E} = 24$ days is a typical value 
for the single lens events 
\citep[Figure 1]{suz16}, and it is usually too short to measure the parallax effect that is 
often measured in events with $t_{\rm E} \simgt 60$ days. Nevertheless, our model includes the parallax effect parameters $\pi_{\rm E} = (\pi_{\rm E, E}, \pi_{\rm E, N})$. The parameters of the best fit 1-planet model are shown in Table \ref{tab:bestfit}.

Note that the best fit 1-planet model shows negative blending ($f_{\rm S} = 
F_{\rm S}/ (F_{\rm S} + F_{\rm B}) > 1$, in Table \ref{tab:bestfit}).
As discussed in many papers \citep{par04,sum06,smi07}, the negative blending does happen 
for combinations of the following reasons: an incorrect light curve model, systematic 
errors in the photometry, and/or indeed, negative blending flux.
The last case occurs when the target happens to be located at a ``hole" in the uneven background 
of unresolved faint stars. The better light curve models discussed later also show 
the negative blending. Furthermore, the amount of the blending, $f_{\rm S} \sim 1.1-1.2$ is quite common \citep{sum06, smi07}, 
and the magnitude of the hole is consistent with unresolved turnoff stars in the Galactic bulge written in later.
Thus, it is unlikely that the incorrect light curve model yields the negative blending.

With a careful visual inspection of the residual plot from the best fit 1-planet model, a weak bump in the MOA data just before the peak magnification was found.
Although this bump could be explained by a systematic scatter, red noise, or other artificial effects, they are unlikely as discussed in Section \ref{sec_real}.
We conduct further light curve analysis in the following sections to examine whether the bump is explained by an additional companion to the lens or source star.

\subsection{The Second Anomaly Has an Astrophysical Origin}
\label{sec_real}

The bump at $\rm HJD' = 6899$ is well explained by an additional planet 
as we will show in the Section \ref{sec_tri}.
Although the triple lens model improves the fit by $\chi^{2}=171$ from the 
1-planet model as written in later, the bump might have been caused by non-astrophysical reasons.
However, this is unlikely for several reasons. 
First, the weather during $6895 < \rm HJD' < 6903$ was continuously stable and clear. The moon phase on $\rm  HJD'=6899$ was 3.5 days after new moon.
Second, we used the light curve using the DoPHOT PSF which is different from the PSF used for the MOA light curve in the main analysis\footnote{The flat and dark images are also different between the MOA light curve in the main analysis and that with the DoPHOT PSF.}, and found the same bump that prefers the triple lens model. Note that we also found the triple lens model using the online MOA data which is less optimized compared to the re-reduced data.
Third, the bump is significantly larger than the variation in the baseline.
We cannot find similar level of bumps and/or dips in the light curve including 10 years baseline except for the first anomaly.
Fourth, we checked the light curves of the stars around the event and found no similar bump at $\rm HJD'=6899$.
From these inspections, we conclude that 
this bump was caused by an astrophysical reason.

%
%
%


\subsection{The Planetary Model Ignoring the First Anomaly}
\label{sec_2pl}
If we assume that the bump at $\rm HJD' = 6899$ is induced by an additional companion to the lens system, the observed light curve should be fitted with a triple lens model \citep{ben99, gau08,ben10,han13,jp16}.
Before a trial of the complicated triple lens model fitting, 
the validity of this assumption can be tested with simplified methods. 
One idea is to use a superposition of two binary lens models, which is 
a good approximation of the triple lens model \citep{han01, rat02, han05}.
The total perturbation by multiple planets from a single lens light curve can be approximated by the sum of perturbations due to each planet. Although this method allows us to estimate the correct triple lens parameters, 
it requires as long computation time as the triple lens model.

Instead, in this event, as the each anomaly occurred within relatively short time scale, and they are well separated in the light curve, we can fit each anomaly separately by ignoring the data points around the other anomaly.
We just remove the data points between $\rm 6884 < HJD' < 6894$ and fit the remaining light curves with a binary lens model, instead of using the superposition of two binary lens models for the full data set.
We follow the same fitting procedure as written in above for the 1-planet model.
We found that a binary lens model with the mass ratio of $q \sim 3 \times 10^{-4}$ (i.e., a planetary model because of $q < 0.03$) and separation of $s \sim 1$ is preferred by $\Delta \chi^{2} \sim 91$ to a single lens model.
Although this significance is just below the detection threshold of planetary signals, $\chi^{2}_{\rm thres} = 100$, 
in the statistical analysis of MOA survey data \citep{suz16}, the reasonable fitting parameters motivated us to try to fit a triple lens model to OGLE-2014-BLG-1722 light curves.

The best planetary model found here have a large source size, $\rho \sim 0.04$ to produce a smooth and weak anomaly by approaching the cusp of a resonant caustic.
However, such a large source size could be ruled out by the data which are intentionally removed in this model fitting.
Since the source trajectory passed the region between the two triangular planetary caustics due to the first planet, 
we would have observed a more significant (longer and possibly larger) anomaly with a hypothetical large source star.
With a smaller source size ($\rho < 0.01$), we found planetary models with $s \sim 0.9$ and $s \sim 1.3$, both of which have almost same $\chi^{2}$ values.
Thus, we expect the possible close-wide degeneracy in the projected separation of the second planet.

\subsection{2-Planet Model}
\label{sec_tri}


A triple lens model requires three additional fitting parameters: the mass ratio of the 
second planet to the host star, $q_{2}\, (\equiv m_{\rm c}/M_{\rm host})$, the projected 
separation between the second planet and host star normalized to the angular 
Einstein radius, $s_{2}$, and the angle between the line connecting the host star 
and first planet and line connecting the host star and second planet, $\psi$.
Thus, the magnification at a given time $t$ is described as 
$A(t) = A(t;t_{0}, t_{\rm E}, u_{0}, q, s, \alpha, \rho, q_{2}, s_{2}, \psi)$. As we 
did for the 1-planet model, MCMC is used to localize the $\chi^2$ minimum for 
the triple lens model. We use the values of the 1-planet model 
parameters for the initial parameters 
for ($t_{0}, t_{\rm E}, u_{0}, \alpha, \rho$). Considering the lens system and 
source trajectory configuration from the 1-planet model and planetary model 
ignoring the first anomaly, we estimate $\psi$ is roughly 2.1 ($120^{\circ}$) and 
use this as an initial value. For the parameters of ($q, s, q_{2}, s_{2}$), we use 
162 combinations of the initial values, which consist of 3, 3, 3 and 6 different 
values for each parameter. Note that the $s_{2}$ covers $0.7 < s_{2} < 1.6$ 
to investigate the possible close-wide degeneracy.


As we see in the 1-planet model,  a significant detection of the parallax effect is unlikely in this event, 
but again we include the parameters, since including the parallax parameters could affect 
the uncertainties of the other parameters \citep{sko16}
and the upper limit could be used as a constraint to estimate the mass of and distance to the lens system.

\begin{figure}
\begin{center}
\epsscale{1.1}
\plottwo{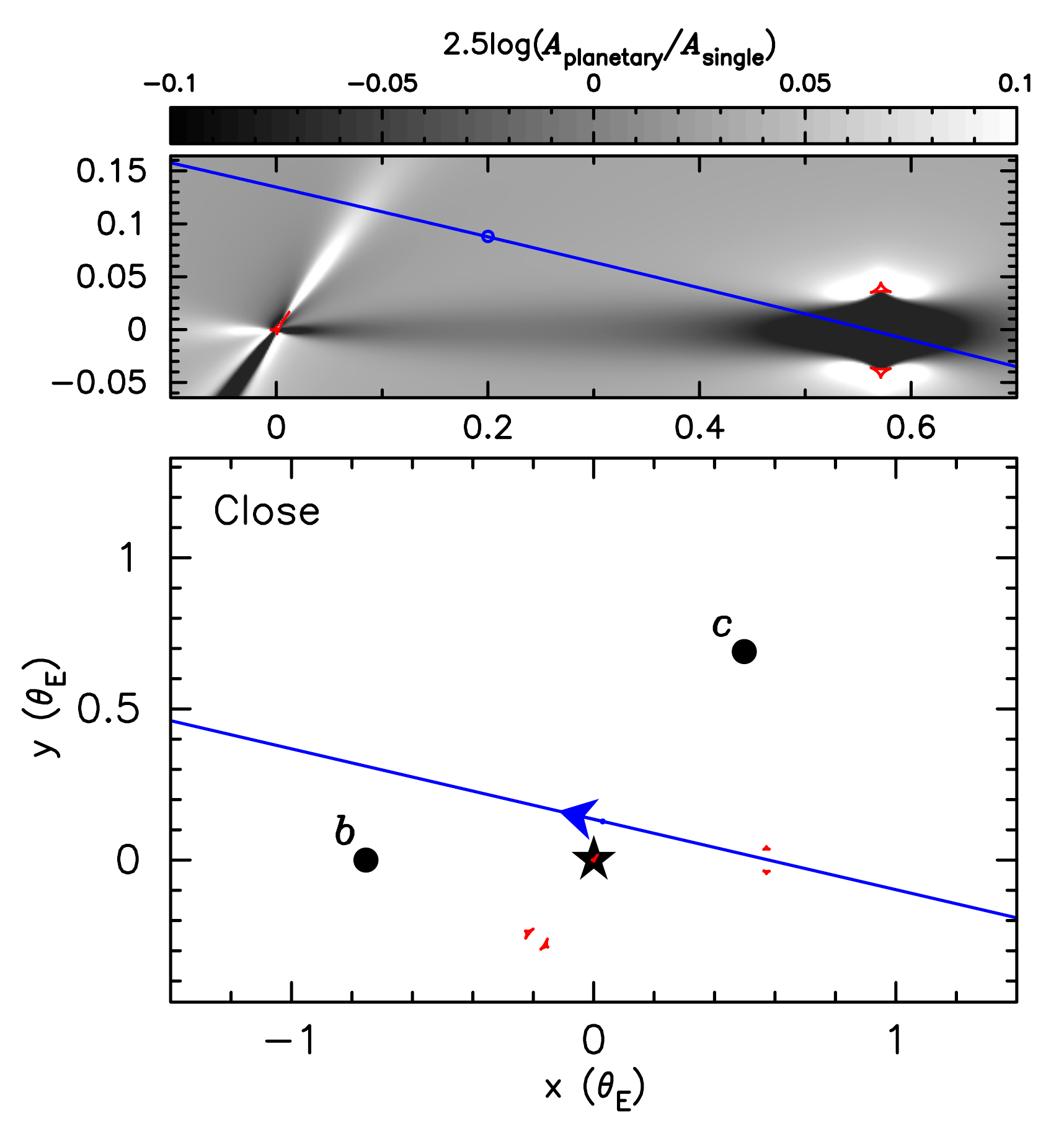}{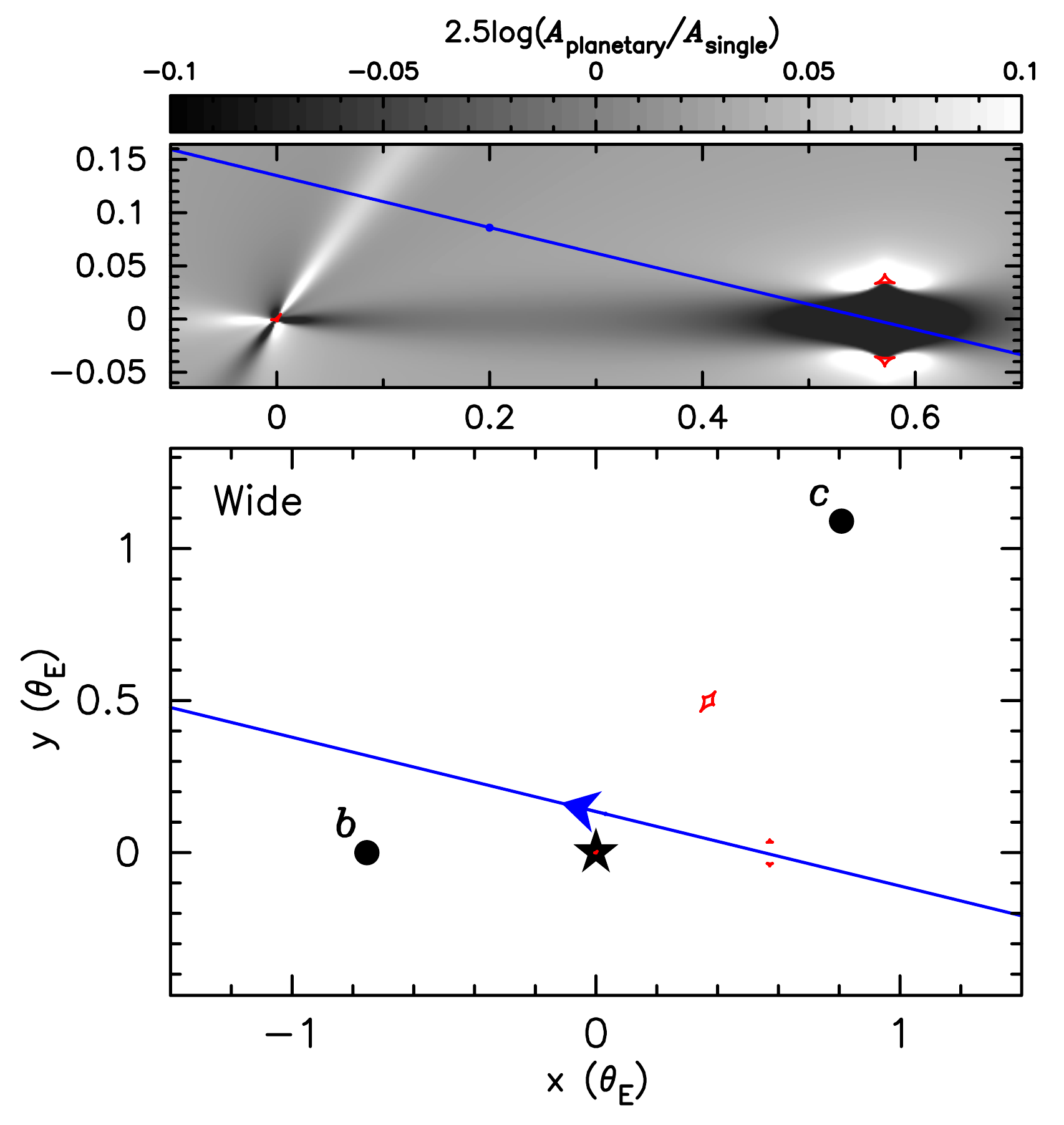}
\caption{The caustic geometry and position of the lens system projected to the sky in unit of the angular Einstein ring radius. The left panels are for close model ($s_{2} < 1$), and the right panels are for wide model ($s_{2} > 1$).
The caustics are drawn in red. The planets and host star positions are indicated with the black circles and star. 
The blue lines and circles indicate the trajectory and size of the source star, which is almost same size as the blue line width, as well as the arrows showing the direction.
The upper panel shows the zoom of the bottom panel, plotted with the relative magnification pattern to the single lens light curve in the gray scale at each sky position. The black and white area have lower and higher magnification than the single lens magnification.}
\label{fig:geometry}
\end{center}
\end{figure}

The best fit 2-planet model is plotted in the Figure \ref{fig:bestfit} and its parameters are shown in Table \ref{tab:bestfit}.
The geometry of the lens system and source trajectory is plotted in Figure \ref{fig:geometry}.
As we expected in the previous subsection, we found close ($s_{2} < 1$) and wide ($s_{2} > 1$) models for the second planet.
The parameters for the two models are consistent with each other, with the exception of $s_{2}$.
The parameters of the primary lens and first planet in this model are also consistent with those of the best 1-planet model.
The $\chi^{2}$ improvement of the best fit 2-planet model (close model) from the best 1-planet model is 171, which is significant if we apply the detection threshold of $\chi^{2}_{\rm thres}=100$.
80\% of the $\chi^{2}$ improvement originates from the 24 data points around $\rm HJD'=6899$, 15\% comes from the data at $\rm HJD'=6900$, and the rest are from the nights around the peak magnification. Only the MOA data contributes to the detection of the second planet, OGLE-2014-BLG-1722Lc.
Note that D.P.B conducted a grid search for the third lens object with all the parameters for the primary star and 1st planet fixed, and found the best model which is consistent with the above best fit 2-planet model.

The amount of the negative blending in the 2-planet model is almost same level as the 1-planet model as written in Table \ref{tab:bestfit}.
The magnitude of the ``hole" is $I_{0, \rm hole} = 19.27 \pm 0.23$ with the extinction correction (See Section \ref{sec_phy}), 
which is consistent with the unresolved turnoff stars in the Galactic bulge.

\begin{figure}
\epsscale{1.0}
\plotone{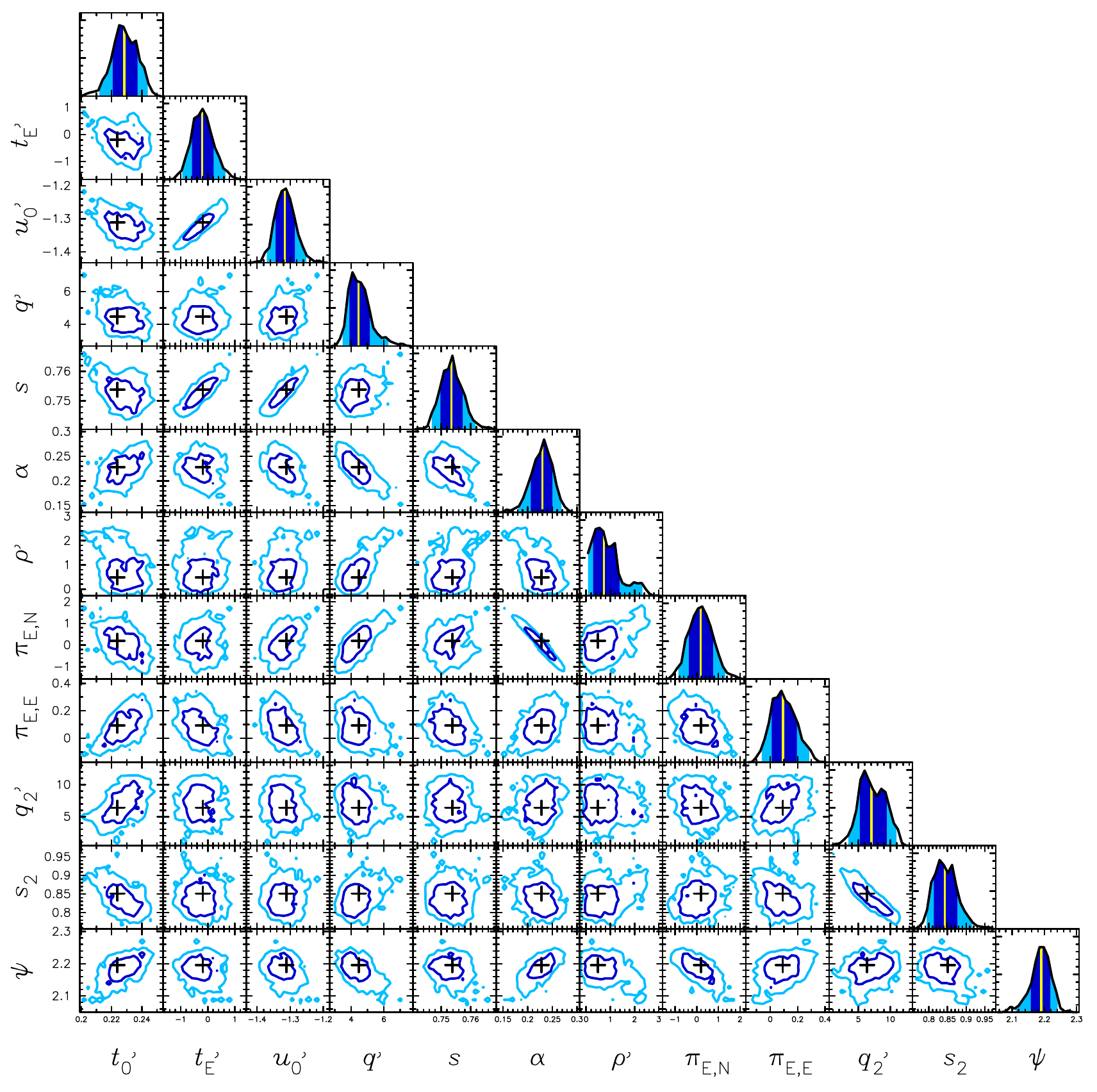}
\caption{Parameter-parameter correlations for the 12 fitted parameters of the close model from the MCMC chain. 
The dark and light blue lines show 1 $\sigma$ and 2 $\sigma$ contours. 
The cross marks show the best fit values in each parameter. In the diagonal panels, 
the distributions are projected to one dimension. The median of each distribution is indicated by the yellow vertical lines. 
The dark and light blue regions correspond to 1 $\sigma$ and 2 $\sigma$ limits on each parameter.
For the visualization reason, some parameters are redefined here: $t_{0}' = t_{0} - 6900$, $t_{\rm E}' = t_{\rm E} + 24$, $u_{0}' = u_{0} \times10$, $q' = q \times 10^{4}$, $\rho' = \rho \times 100$, and $q_{2}' = q_{2} \times 10^{4}$.}
\label{fig:chain}
\end{figure}

Figure \ref{fig:chain} shows parameter correlations for the MCMC chain for the best fit close model. 
The mass ratio of planet b, $q$, is measured with 16\% uncertainty, 
but the mass ratio for planet c, $q_{2}$, has a large uncertainty of $\sim 33\%$. 
This is simply because the anomaly of planet c in the light curve is relatively weak. Nevertheless, the projected separations of both planets are well measured, as the separation is much sensitive to size of caustics compared to the mass ratio \citep{chu05,han06}.
Unlike the previous two-planet microlensing events \citep{gau08, han13}, the relative source star size and the Earth's orbital parallax effect are not well constrained, because the source trajectory did not hit the caustics and/or cusp, and the event time scale is not long enough to measure the parallax. 
Thus, we can estimate the mass and distance of the lens system only from a Bayesian analysis based on a standard Galactic model and the assumption that the planet hosting probably does not depend on the mass and distance of the lens star \citep{jp06,suz14}.


\subsection{Alternative Models}
\subsubsection{Stellar Companion to the Lens}

It is unlikely that a stellar companion to the lens (i.e., a circumbinary 
planetary system or a planet orbiting one member of a binary system) 
causes the second anomaly, because the stellar binary solutions were 
ruled out by $\Delta \chi^{2} > 220$ when we conducted the binary lens 
modeling excluding the data points around the first anomaly.

\subsubsection{Binary Source Model}
A binary source model is also a possible alternative explanation of the second anomaly \citep{gau98_BS,jun17bs}.
In this event, a hypothetical second source star is expected to 
be very faint and highly magnified, since the apparent duration of the
anomaly is very short ($\sim 1\,\rm day$) compared to 
that of the primary event ($\sim 1\, \rm month$), and we do not 
see any excess flux when the primary source was demagnified 
(compared to the single lens--single source model) at the first anomaly.
To test this binary source scenario, we fit the light curves with a binary source model 
(more specifically, a binary lens plus binary source model), which is also 
used for the modeling of MOA-2010-BLG-117, the first\footnote{First case, where the both source stars were magnified.} planetary microlensing event with a binary source \citep{ben17}, and OGLE-2016-BLG-1003, the first resolved caustic-crossing binary source event \citep{jun17blbs}.
Compared to a binary lens model with a single source star, the 
binary source model requires additional three parameters: the 
time of the peak magnification for the second source, $t_{0, 2}$, 
the impact parameter of the second source divided by the Einstein 
radius, $u_{0, 2}$, and the flux ratio of the second source 
to the primary source, $f_{\rm ratio}$.
We found that the best fit binary source model is also able to explain the small bump.
The best fit parameters are given in Table \ref{table-BS} with 
the apparent brightness of the source stars and blending parameter as well.
Note that there is a 2-fold degeneracy in $u_{0,2}$: the positive and negative 
value in $u_{0,2}$ shows the same magnification.
These degenerate models yields different values of projected separation between the two sources.
The amount of the blending is also very similar to that of 1-planet and 2-planet models.
The best fit binary source model is slightly disfavored compared to 
the best 2-planet model by $\Delta \chi^{2}$ of 4.2 ($2.0\, \sigma$).
Note that we use the static model (no microlensing parallax parameters) for the $\chi^2$ comparison, since the 2-planet static model and the 1-planet binary source model have the same number of modeling parameters.
Thus, we cannot simply rule out the binary source scenario from the light curve fitting.

However, the occurrence rate of the stellar companion to explain the 
best binary source model is not very high.
Assuming the projected Einstein radius on the source plane, 
$\hat{r}_{\rm E} = 2.93\, \rm AU$ that is estimated using 
the Galactic model \citep{han95} with a constraint of $t_{\rm E}$, 
the projected physical separation of the binary source is 
$0.40_{-0.15}^{+0.20}$ and $0.44_{-0.17}^{+0.21}\,\rm AU$ for the positive and negative $u_{0,2}$, respectively.
Although we do not have color information for the primary source star, 
it is safe to assume that the primary is a solar type star.
With an assumption that the period distribution of nearby solar type 
stars \citep{rag10} is applicable to the stars in the Galactic bulge, 
the estimated projected physical separation between the two sources 
corresponds to the tail of the period distribution of companion stars in \citet{rag10}.
To fairly compare the relative probability of the 2-planet and binary source scenario, 
we consider the $\chi^{2}$ difference, the prior probability from the mass and 
separation distribution, and the detection efficiency as well.

For the binary source scenario, we use the companion distribution of \citet{rag10} to estimate the prior probability.
For the 2-planet scenario, the mass ratio function of \citet{suz16} is used.
The relative prior probability we obtained is 
$0.041$ and $0.043$ for the positive and negative $u_{0,2}$ binary source model, and 
$0.152$ and $0.192$ for the close and wide $s_{2}$ 2-planet models.
The detection efficiency we obtained is 
$0.098$ and $0.052$ for the positive and negative $u_{0,2}$ binary source, and 
$0.46$  and $0.12$ for the close and wide $s_{2}$ 2-planet models, respectively.
With the above values, we found that the relative probability of 
0.991 and 0.009 for the 2-planet and binary source scenario, respectively.
The binary source scenario is marginally ruled out by $3.1\,\sigma$, if we consider that the relative probability is equivalent with the weight of $e^{-\Delta \chi^{2}/2}$.

%


%


%
%
%

\section{Estimate of the Physical Parameters}
\label{sec_phy}

The microlensing method allows us to measure the mass of and distance to the lens system when we measure the microlenisng parallax effect and finite source effect \citep{gou00}. 
It is also possible to measure the mass and distance of the lens if we can determine 
the flux of the lens star in addition to a measurement of microlensing parallax or 
the finite source effect \citep{ben10,ben15hst, bat15}. 
But, it is important to verify that the excess flux at the position of the source 
is really due to the lens \citep{apa17, kos17_contami}.
Unfortunately, OGLE-2014-BLG-1722 showed neither the parallax effect nor finite source effect in the observed light curve. 
Thus, we need to use the Bayesian approach to estimate the mass of the lens system. 
As a prior, we use the Galactic model of \citet{han95} and the mass function of \citet{sum11}.
The constraints we use here to obtain the posterior distribution of the lens mass 
and distance are the observed $t_{\rm E}$ and upper limit of $\pi_{\rm E}$.

\begin{figure}
\epsscale{0.65}
\plotone{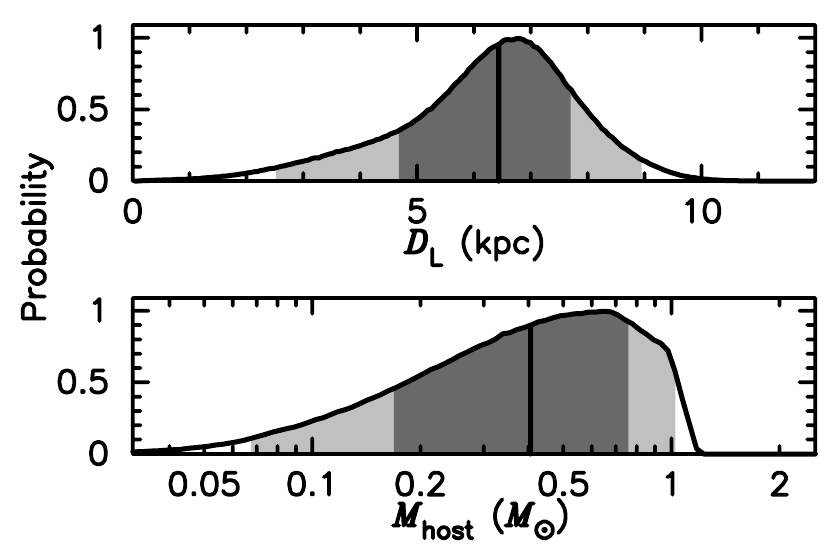}
\caption{Probability distribution from the Bayesian analysis for the distance to and mass of the host star.
The dark and light gray regions show the $68\, \%$ and $95\,\%$ confidence intervals, and the vertical black lines indicate the median value.}
\label{fig:phyhost}
\end{figure}

\begin{figure}
\plotone{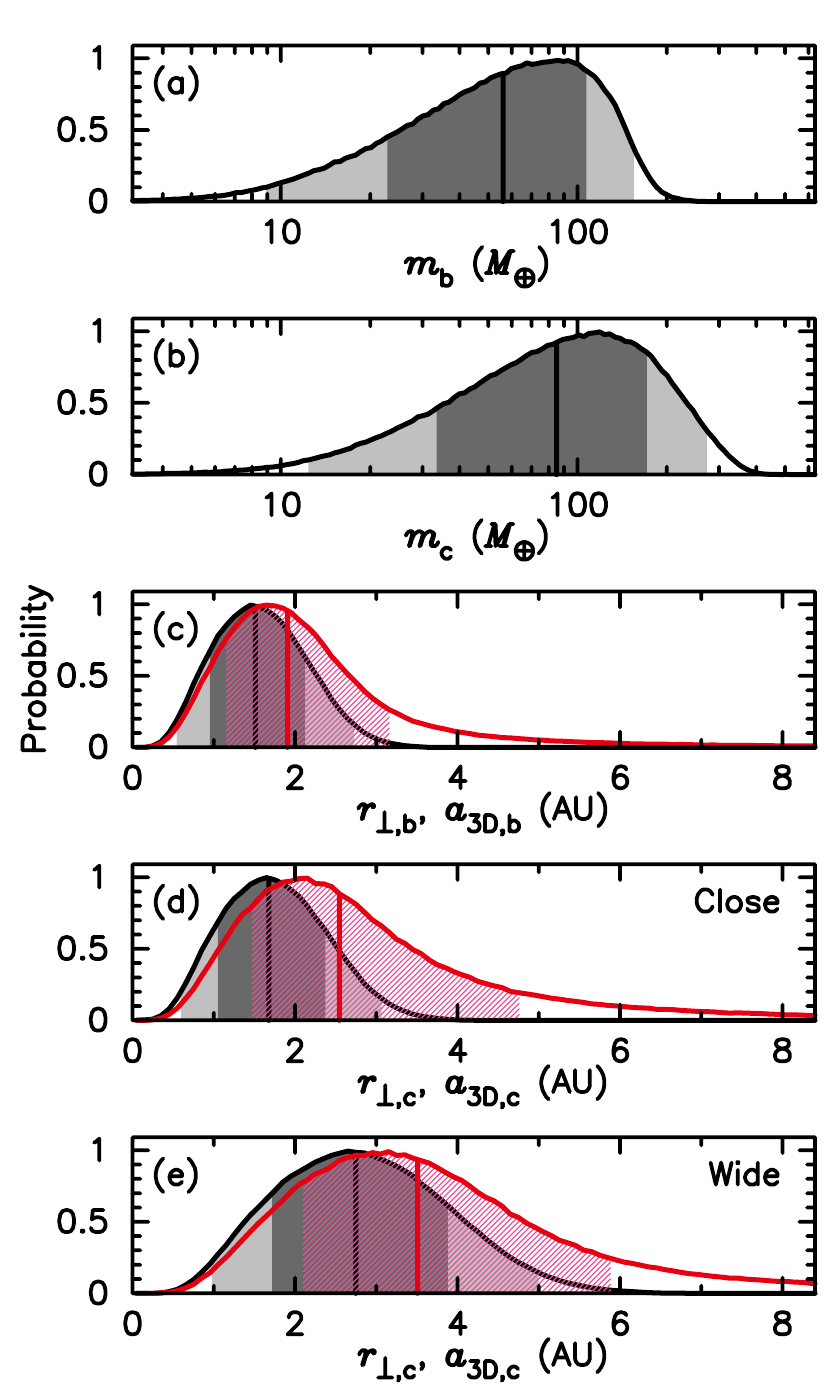}
\caption{Same as Figure \ref{fig:phyhost}, but for the physical parameters of planet b and c. (a) and (b) indicate the mass of each planet. (c), (d) and (e) indicate the projected separation (in black) and 3-D separation (in red) with the assumption of a circular orbit and random orientation for planet b, and close and wide model of planet c, respectively. For the 3-D separation, only $68\, \%$ confidence intervals are plotted with the red-shaded area.
}
\label{fig:phyplanet}
\end{figure}

The estimated physical values of the lens system are written in Table \ref{tab:physical}. 
As we expect, 
the physical parameters of the close and wide solutions are similar each other except for the projected/deprojected separation between the host and planet c (i.e., $r_{\perp, \rm c}$ and $a_{\rm 3D, c}$).
Thus, we combine the posterior distribution of close and wide models with a weight of $e^{-\Delta \chi^{2}/2}$, where $\Delta \chi^{2}$ is the $\chi^{2}$ difference of the two models, $\Delta \chi^{2} = 1.2$, and report the combined values of each parameter except for $r_{\perp, c}$ and $a_{\rm 3D, c}$. 
The posterior distribution of the distance to the lens system and mass of the host star are shown in Figure \ref{fig:phyhost}. 
We found that the host star mass is $M_{\rm host} = 0.40_{-0.24}^{+0.36} \,M_{\odot}$ and 
the lens system is located at $D_{\rm L} = 6.4_{-1.8}^{+1.3} \,\rm kpc$ from our solar system.
Assuming that the planet frequency of a system like OGLE-2014-BLG-1722Lb,c 
is uniform in the host star mass and distance, we use the 
distribution of the host star mass and distance to estimate 
the physical values of the planets. 
Figure \ref{fig:phyplanet} shows the posterior distribution 
of the mass and (projected and deprojected) separation of the two planets.
We found that the mass of planet b and c are 
$m_{\rm b} = 56_{-33}^{+51} \,M_{\oplus}$ and 
$m_{\rm c} = 85_{-51}^{+86} \,M_{\oplus}$.
The projected separation between the host and planet b is $r_{\perp, \rm b} = 1.5 \pm 0.6\,\rm AU$.
For planet c, the projected separation from the host is 
$r_{\perp, \rm c} = 1.7_{-0.6}^{+0.7} \,\rm AU$ and 
$r_{\perp, \rm c} = 2.7_{-1.0}^{+1.1} \,\rm AU$ for the close and wide models, respectively.

Revealing semi-major axis, inclination and eccentricity of the two cold planets can 
provide stronger constraints on the formation and evolution of this system.  
The Jupiter-Saturn analog system, OGLE-2006-BLG-109, provided us its orbital information 
since the effects of the orbital motion in the lens system were observed \citep{ben10}.
For OGLE-2014-BLG-1722, however, such a constraint cannot be obtained because we did 
not detect the orbital motion effect of the planets due to the short planetary perturbations and no caustic crossings.
We assume co-planar and circular orbits to estimate the semi-major axis of the planets.
Also, we apply the Hill stability constraint on planet c \citep{gla93} with a given $a_{\rm 3D, b}$.
With the random inclination and orbital phase, we estimated that semi-major axis of planet b is $a_{\rm 3D, b} = 1.9_{-0.8}^{+1.3}\, \rm AU$.
The estimated semi-major axis of planet c is $a_{\rm 3D, c} = 2.5_{-1.1}^{+2.2}\,\rm AU$ 
and $a_{\rm 3D, c} = 3.5_{-1.4}^{+2.4}\,\rm AU$ for the close and wide models, respectively.
Assuming the snow line of $a_{\rm snow} = 2.7 M_{\rm host}/M_{\odot} \,\rm AU$ \citep{kk08}, 
both planets are likely located at 1.4--2.4 times the snow line, which is typical for microlens planets.


%

\begin{figure}
\plotone{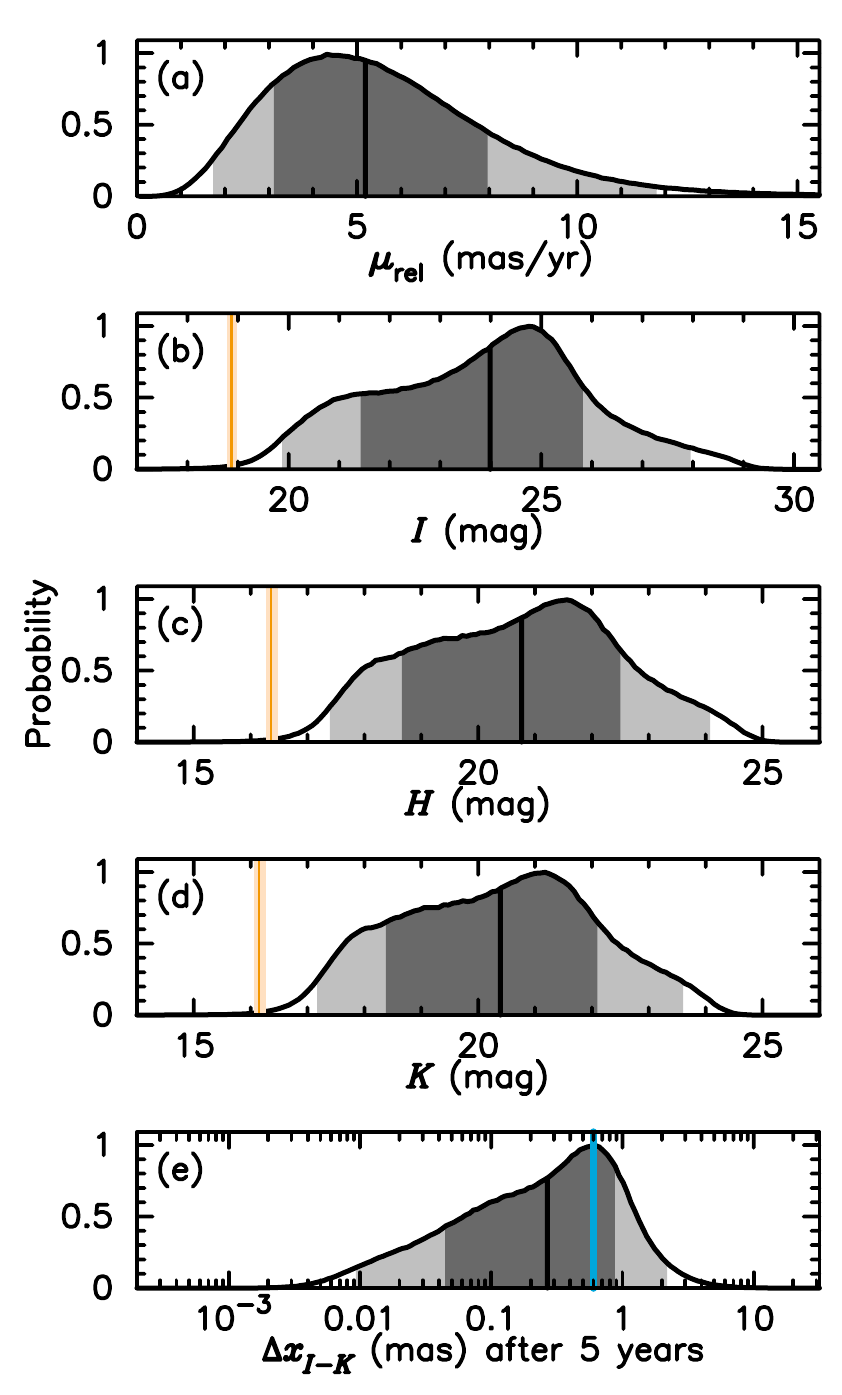}
\caption{Same as Figure \ref{fig:phyhost}, but for the relative proper motion in geocentric frame and lens brightness with extinction. (b), (c) and (d) indicate the lens brightness in $I$, $H$ and $K$, as well as the source star magnitude in the orange vertical line with $3\,\sigma$ uncertainty.
The source star magnitude in $H$ and $K$ are estimated using the PARSEC isochrones, version1.2S \citep{bre12,che14,che15,tan14}.
(e) shows the color dependent centroid shift in 5 years for $I$ and $K$-band.
The blue vertical line shows $0.6\, \rm mas$, which was actually achieved for OGLE-2003-BLG-235 \citep{ben06} with {\it HST}. 
{\it JWST} would measure the color dependent centroid shift with even better 
accuracy using F070W or F090W, and F200W filters, which have wavelength coverage roughly similar to $I$ and $K$-band.}
\label{fig:lensmag}
\end{figure}

Figure \ref{fig:lensmag} shows the expected apparent lens brightness in {\it I}, {\it H} and
 {\it K}-bands, as well as the expected relative proper motion, $\mu_{\rm rel}$ in the geocentric frame. 
These distributions are derived based on the Bayesian analysis described above with 
a constraint of the observed $t_{\rm E}$ and upper limit on $\pi_{\rm E}$.
The extinction in $I$-band to the source star, $A_{\rm I, S}=1.75 \pm 0.09$, is estimated by comparing 
the apparent position of Red Clump Giant (RCG) on the OGLE III CMD, 
$(V-I, I)_{\rm RCG} = (2.650, 16.258) \pm (0.014, 0.003)$, 
with the intrinsic value $(V-I, I)_{\rm 0} = (1.06, 14.51) \pm (0.06, 0.09)$ from \citet{bens11} and \citet{nat13}. 
The extinction in $H$ and $K$-band to the source was estimated by using the BEAM calculator \citep{gon12}, where we take an average for the extinction law of \citet{car89} and \citet{nis09}.
Then, following \citet{ben15hst}, the extinction to the foreground lens is given by
\begin{equation}
A_{\Lambda, \rm L} = \frac{ 1-e^{-D_{\rm L}/h_{\rm dust}} } { 1-e^{ -D_{\rm S}/h_{\rm dust} } } A_{\Lambda, \rm S},
\end{equation}
where the index $\Lambda$ refers to the passband: $I$, $H$ or $K$, and $h_{\rm dust} = (0.10 \pm 0.02\,{\rm kpc})/{\rm sin}\,b$.
Unfortunately, the expected lens brightness is much fainter than 
the source star, even for the {\it K}-band and thus it will be difficult 
to detect the lens flux on the superposition of the relatively bright source.
Instead, the color-dependent centroid shift \citep{ben06,ben07,ben15hst,apa17} 
in $I$ and $K$-band, $\Delta x_{I-K}$, would be detectable in 5 years 
as shown in the bottom panel in Figure \ref{fig:lensmag}.
\citet{ben06} detected $\Delta x_{V-I}$ of $0.6\, \rm mas$ for OGLE-2003-BLG-235 using {\it Hubble Space Telescope} ({\it HST}).
We expect that {\it James Webb Space Telescope} ({\it JWST}) would measure this effect with even better accuracy and put much tighter constraint on the physical parameters of the lens system.

\section{Occurrence Rate of Multiple Gas Giant Planet Systems}
\label{sec_occ}

The light curve of this event was well observed by the survey teams. 
Especially, MOA had dense coverage each night while the source star was magnified, 
which resulted in higher detection efficiencies to planets than the average efficiency of MOA survey in 2007-2012 \citep{suz16}.
Therefore, an extended study of the statistical analysis of the MOA data will automatically include this event.
Here, we can roughly (but robustly) estimate the occurrence rate of multiple cold giant planet systems by using the $\mu$FUN \citep{gou10} and an extended MOA survey data, and the two detections of such systems: OGLE-2006-BLG-109Lbc and OGLE-2014-BLG-1722Lbc.
We can constrain the occurrence rate of the multiple systems only for cold giant planets from the current microlensing data for the following reasons. 
First, the most sensitive region of the microlensing method is located at a few times of the snow line. 
Second, the detection efficiency of two planet systems with masses of Neptune mass (or below) is too low to put a strong constraint on its occurrence rate (because the sensitivity is approximately the product of detection efficiency for each planet).
Third, the detected multiple planet systems included in this analysis are Jupiter-Saturn analogue and two-Saturn system.
More specific definition of multiple cold giant planet systems will be described later in this section.

To estimate the occurrence rate, we need to consider the detection 
efficiency \citep{gou10,cas12,shv16,suz16} of both planets in each observed event.
First, we assume that the detection efficiency of two-planet systems can be approximated as $\epsilon_{12} \equiv \epsilon(q_{1}) \times \epsilon(q_{2})$. 
This is safe especially for the sample dominated by low magnification events \citep{gau98_multi,hanpark02} like the MOA survey data.
Second, we assume that the MOA's survey sensitivity (the total detection efficiency) in 2013-2016 is similar to that of 2007-2012. 
We randomly pick up four years from 2007-2012 and use the detection efficiency in each event as the hypothetical MOA data of 2013-2016 (The full analysis of 2013-2016 data is beyond the scope of this paper.). 
Then, one event randomly selected in the hypothetical 2014 data is substituted by the real detection efficiency of OGLE-2014-BLG-1722 (whose MOA name is MOA-2014-BLG-490).
We repeated this process and confirmed that the occurrence rate does not depend on the data used for the hypothetical survey data.
For the $\mu$FUN sample, we use the detection efficiencies estimated by \cite{gou10}. For the MOA sample, we use the detection efficiencies in \cite{suz16}, which are determined by using the method of \cite{rhi00} with an 
assumption of logarithmically uniform planet distribution in both mass ratio and separation.
Thus, the total sample we use here consists of 13 events from $\mu$FUN (2005-2008) and about 2400 events from MOA (2007-2016).
The overlapping events are removed from the MOA sample, and we use the $\mu$FUN detection efficiencies for them.

The occurrence rate of multiple cold giant planet systems, $\eta_{\rm multi}$, can be estimated by maximizing the Likelihood,
\begin{equation}
L(\eta_{\rm multi}) \equiv P(\eta_{\rm multi}|d) \propto P(\eta_{\rm multi})\prod_{i}^{N} f_{i} \prod_{j}^{M-N} (1-f_{j}),
\end{equation}
where $P(\eta_{\rm multi}|d)$ shows the probability of $\eta_{\rm multi}$ conditioned on the data $d$, $N$ and $M$ are the number of detected multiple systems and all the events included, respectively, and $f$ is the product of detection efficiency and the occurrence rate: $f = \epsilon_{12} \times \eta_{\rm multi}$. We assume that $P(\eta_{\rm multi})$ is a uniform distribution, 
thus the Likelihood function is given by the binomial distribution. 
For each event, we integrate the detection efficiency within $-0.5 < {\rm log}\,s < 0.5$ where the sensitivity is highest 
and, use the mass ratio range of $5 \times 10^{-5} < q < 0.03$ to compute $\epsilon_{12}$.
As indicated in Figure \ref{fig:occ_cl90}, we use a box that has $0.5\,\rm dex$ in each side on the $q_{1}$--$q_{2}$ plane, and calculate the average detection efficiency in this box. 
The occurrence rate at the center of the box is calculated by using the averaged detection efficiency. Then, by moving the box on this plane, we calculate the occurrence rate of the system with any combination of $q_{1}$ and $q_{2}$, where $q_{2} > q_{1}$.
Thus, $\eta_{\rm multi}$ is the occurrence rate for the planetary systems whose host stars are orbited by at least two planets both of which are located in the separation of $-0.5 < {\rm log}\,s < 0.5$ and in a 0.5 dex bin centered on ${\rm log}\,q_{j}$, where $j$ is 1 or 2.

We computed the upper limit on the occurrence rate at the 90\% confidence level, which is indicated in Figure \ref{fig:occ_cl90}.
As we expect from the detection efficiency, the right upper corner in the figure has 
a lower occurrence rate limit and left bottom corner has a higher occurrence rate limit.
We averaged over the upper limit in the area where the box can include the two detections, and found the $\eta_{\rm multi} < 0.25$ with a 90\% confidence level.
Furthermore, we estimate the mode, median and $1\,\sigma$ uncertainty of the occurrence rate in 
the same parameter space. 
As indicated in Figure \ref{fig:occ_mode}, we found $\eta_{\rm multi} = 0.12_{-0.06}^{+0.09}$, with a mode of $0.09$.
This result is not sensitive to the size of the sliding box.
Also, note that the mass ratios of Jupiter and Saturn in our solar system happen to be included in the area where we calculate the $\eta_{\rm multi}$.


The most likely value of the occurrence rate, $\eta_{\rm multi} = 0.09$ can be compared with the frequency of the solar-like systems estimated by \citet{gou10}, 0.17, which is consistent with $\eta_{\rm multi}$ due to the large uncertainty, but factor 1.9 larger than our mode value.
They considered scaled solar system including Uranus and 
Neptune\footnote{Neptune is just outside of ${\rm log}\,s=0.5$, 
if we assume that the Einstein radius ($s=1$) of a solar-like analogue corresponds to $8\,\rm AU$} analogues, 
so the definition of the solar-like system is different from that of multiple cold gas giant planet system.
In their simulation, however, most of the detected multiple systems are Jupiter-Saturn analogues.
Also, their estimated number (0.17) would decrease by about factor two if we consider an extended analysis of \citet{gou10}. 
They found 6 planets with 1 multiple system in their 4 years of $\mu$FUN data, but we expect that only 3 planets \citep{bac12,fuk15,yee12} (and no multiple system) would be included in the next 6 years.
Note that OGLE-2012-BLG-0026 \citep{han13} would not be included in the simple extended study, because it's peak magnification does not satisfy the $u_{0} < 0.005$ criterion in \citet{gou10}.
Therefore, with the assumption of a linear extrapolation of the 4 year sensitivity to 10 years, the frequency of solar-like system would be roughly 0.07, which is closer to our mode value.

Considering that the average number of cold planets with the mass ratio of $q=4.5\times10^{-4}$ per star is 0.14, which is estimated by integrating the best fit mass ratio function of \citet{suz16} within the separation of $-0.5 < {\rm log}\,s < +0.5$ and mass ratio of $  -3.6 < {\rm log}\,q < -3.1$, the estimated mode of $\eta_{\rm multi}$ could be overestimated.
If we assume that every planetary systems with cold gas giants are 2-planet systems, the occurrence rate of 2-planet system will be 0.07 at maximum. Thus, we can use the average number of planets for a prior probability of the upper limit on $\eta_{\rm multi}$.
We use the uncertainty in the mass ratio function in \citet{suz16} to estimate the upper limit distribution on $\eta_{\rm multi}$, whose median and 1\,$\sigma$ uncertainty is 0.058  and 0.008, respectively.
Here, we assume that the upper limit on $\eta_{\rm multi}$ of the systems with $q_{1} = q_{2} = 4.5 \times 10^{-4}$ is more conservative than that of the systems with $q_{1} = 4.5 \times 10^{-4}$ and $q_{2} = 3.0\times 10^{-3}$.
Adopting 3\,$\sigma$ uncertainty for the prior distribution, we estimate $\eta_{\rm multi} = 0.06 \pm 0.02$, with the mode of 0.07.
The estimated occurrence rate is strongly depending on the assumed prior distribution of $\eta_{\rm multi}$. Nevertheless, the constrained $\eta_{\rm multi}$ is more consistent with the hypothetically extended study of \citet{gou10} in the previous paragraph.
Although there are still large uncertainties in $\eta_{\rm multi}$, future studies of the occurrence rate of the multiple cold planet systems are important to understand the planet formation beyond the snow line.

%
%
%
%

\section{Summary}
\label{sec_sum}

The microlensing event OGLE-2014-BLG-1722 showed two relatively short deviations from the expected single lens light curve.
The triple lens model consisting of a primary lens star and two planetary mass ratio objects is the best fit model to explain the two anomalies.
Thus, if the 2-planet model is true, this is the first detection of two-planet system in the low magnification microlensing events, 
and the third detection of two-planet system found by the microlensing method.
Also, this event is the first two-planet system detected in pure survey data.
We found that the first anomaly was caused by planet b with a mass ratio 
of $q = (4.5_{-0.6}^{+0.7}) \times 10^{-4}$ and separation of $s = 0.753 \pm 0.004$. 
The second anomaly revealed planet c by the $\chi^{2}$ improvement of $\Delta \chi^{2} \sim 170$ compared to the 1-planet model, but its separation has degenerate close-wide solutions.
For the close solution, the mass ratio and separation for planet c are $q_{2} = (7.0_{-1.7}^{+2.3}) \times 10^{-4}$ and $s_{2} = 0.84 \pm 0.03$, respectively.
For the wide solution, they are $q_{2} = (7.2_{-1.7}^{+1.8}) \times 10^{-4}$ and $s_{2} = 1.37 \pm 0.04$.

The physical properties of the lens system could not be measured 
because neither finite source nor microlensing parallax effects were measured.
Instead, we estimated the mass of and distance to the lens system with 
a Bayesian analysis using a standard Galactic model assuming that the planet hosting probability does not depend on host star mass and distance. 
The estimated host star mass and distance are 
$M_{\rm host} = 0.40_{-0.24}^{+0.36}\,M_{\odot}$ and 
$D_{\rm L} = 6.4_{-1.8}^{+1.3}\,\rm kpc$, respectively.
We estimate that the mass of planet b and c are 
$m_{\rm b} = 56_{-33}^{+51}\,M_{\oplus}$ and $m_{\rm c} = 85_{-51}^{+86}\,M_{\oplus}$.
The projected distance between the host star and planet b is $r_{\perp, \rm b} = 1.5 \pm 0.6\, \rm AU$. 
For close and wide models of planet c, the projected separations are 
$r_{\perp, \rm c} = 1.7_{-0.6}^{+0.7}\, \rm AU$ and $r_{\perp, \rm c} = 2.7_{-1.0}^{+1.1}\, \rm AU$, respectively.
We also estimated the semi-major axis of the planets assuming circular and coplanar orbits. 
The semi-major axis of planet b is $a_{\rm 3D, b} = 1.9_{-0.8}^{+1.3}\, \rm AU$, and 
that of planet c is $a_{\rm 3D, c} = 2.5_{-1.1}^{+2.2}\,\rm AU$ and $a_{\rm 3D, c} = 3.5_{-1.4}^{+2.4}\,\rm AU$ for close and wide models, respectively.
Therefore, OGLE-2014-BLG-1722L is a late type star orbited by two Saturn-mass planets that are 
located at a separation of a few times of the snow line.

%
%

%

Although the 2-planet scenario is preferred by $3.1\, \sigma$, one cannot rule out the interpretation where the source is a binary.
For OGLE-2005-BLG-390 \citep{jp06} and OGLE-2016-BLG-1195 \citep{bon17}, a binary source model was ruled out by $\Delta \chi^{2}=46$ and $\Delta \chi^{2}=120$, respectively.
Nevertheless, the marginal detection of the planet c is important for future statistical studies.
The statistical study of 6 year MOA data \citep{suz16} dealt with an ambiguous event where the planetary and stellar binary models comparably well explain the observed light curves. 

Generally, triple lens events are challenging to find the best model because of 
not only the vast parameter space we need to explore but also possible degenerate models. 
The lens system of a circumbinary planet event, OGLE-2007-BLG-349 \citep{ben16} was initially thought to be a two-planet system.
Also, the one-planet event with a binary source, MOA-2010-BLG-117 \citep{ben17} initially mimicked a circumbinary planet event.
Sometimes, the orbital motion of lens system could produce an anomaly in the light curve, which apparently seems to be a signal from the third body \citep{ben99,alb00,uda15,han16}.
In these cases, the possible degenerate solutions were finally ruled out, but there is one triple lens event waiting for modeling.

Although the detection efficiency of multiple planets are very high in high magnification events \citep{gau98_multi}, 
the separations could suffer from the close-wide degeneracy.
[OGLE-2006-BLG-109Lbc \citep{gau08} was detected through the high magnification channel, but the degenerate models were ruled out nonetheless.]
On the other hand, low magnification events are generally expected to have 
less degenerate models and have unique solutions. 
However, we need to overcome the low detection efficiency by continuous light curve coverages.

The two two-planet systems previously detected by microlensing, OGLE-2006-BLG-109L and OGLE-2012-BLG-0026L, were well characterized because both finite source and parallax effects were measured, and the follow-up high-resolution images put further constraints on the lens physical properties.
For most of the planetary events including two-planet systems, however, we expect low magnification with a moderate event time scale like OGLE-2014-BLG-1722, where the mass measurement is difficult from the ground-based telescopes alone.
Future space-based microlensing surveys, such as $\it{WFIRST}$ \citep{spe15} and hopefully $\it{Euclid}$ \citep{pen13} as well, are expected to measure the physical parameters of the lens systems for most of the observed microlensing events including two-planet events, which will eventually allow us to study the occurrence rate of multiple cold planet systems as a function of host mass and Galactocentric distance.

\begin{figure}
\epsscale{0.85}
\plotone{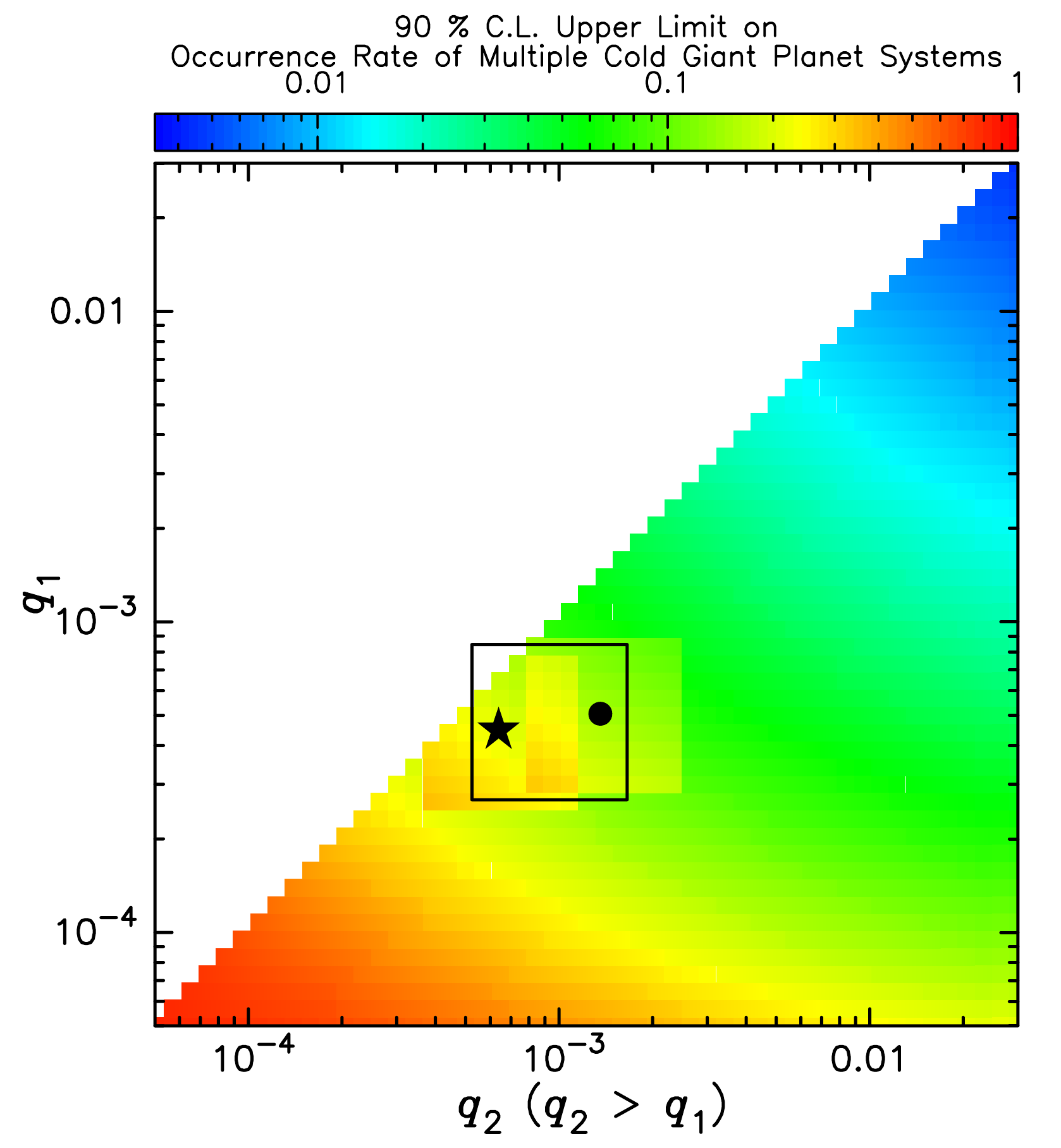}
\caption{$90\,\%$ confidence level upper limit on the occurrence rate of multiple cold giant planet systems.
The occurrence rate are calculated with a moving $0.5\,\rm dex \times 0.5\,dex$ box whose size is indicated at the arbitral position. The black circle and star indicate the mass ratios of two planets in 
OGLE-2006-BLG-109L and OGLE-2014-BLG-1722L systems, respectively. 
Note that the mass ratios of Jupiter and Saturn to the Sun, i.e., 
$q_{2} = 9.5 \times 10^{-4}$ and $q_{1} = 2.9 \times 10^{-4}$ are 
also located within the box when it is placed to include the two detected systems.}
\label{fig:occ_cl90}
\end{figure}

\begin{figure}
\begin{center}
\plotone{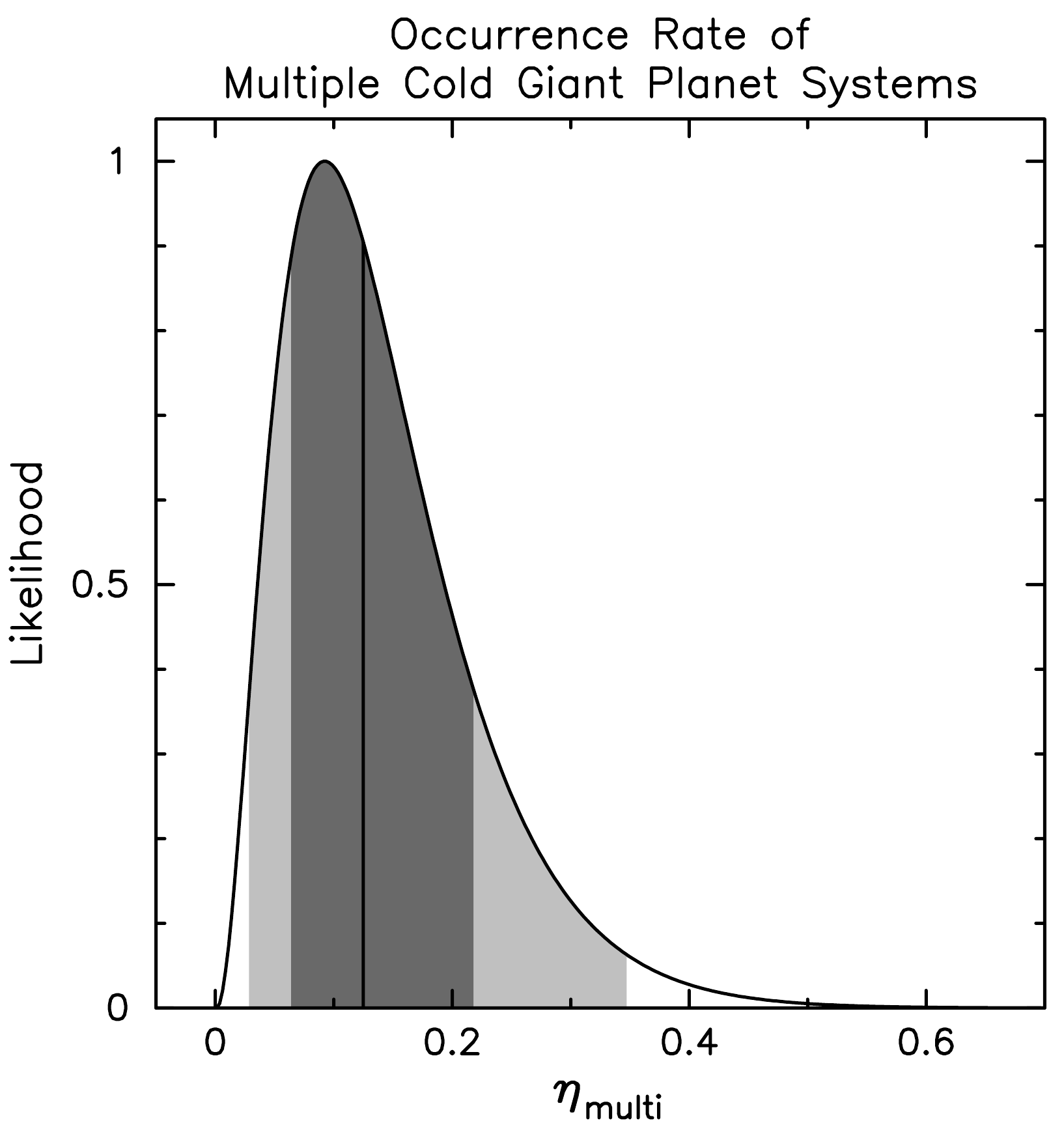}
\caption{Probability distribution for the occurrence rate of cold giant planet systems, which are 
similar to OGLE-2006-BLG-109Lbc and OGLE-2014-BLG-1722Lbc. 
The box for calculating the occurrence rate was used to include the two systems, and we take average the occurrence rate, inside the box.
}
\label{fig:occ_mode}
\end{center}
\end{figure}

The occurrence rate of multiple cold giant system $\eta_{\rm multi}$ were estimated in Section \ref{sec_occ} with the two detections of such systems (OGLE-2006-BLG109Lbc and OGLE-2014-BLG-1722Lbc) in the sample of the 4 year $\mu$FUN data \citep{gou10} and 10 year MOA survey data, which consists of the 6 year full analysis of detection efficiency \citep{suz16} and the simulated hypothetical 4 year data.
We found $\eta_{\rm multi} = 0.12_{-0.06}^{+0.09}$ with the mode value of 0.09.
Thus, roughly 10\% of stars, which specifically could become a lens star of microlensing events toward the Galactic bulge, are orbited by two cold giant planets.
To confirm this result, we need larger statistics of multiple planet system detections.
The occurrence rate of 90\% confidence level upper limit (Figure \ref{fig:occ_cl90}) shows lower and higher occurrence rate for massive and less massive planet pairs.
This can be inferred because massive planets easily affect the orbits of other planets and they might let the other planets migrate to in/outward.
Also, we know that there are many tightly packed coplanar systems consisting of super-Earth or Earth mass planets found by {\it Kepler}.
Thus, we would speculate that multiple cold Neptunes and/or super-Earth systems could be common. 
Such systems can be found only by microlensing survey. 
The current ground-based microlensing surveys do have sensitivities to such systems, but the sensitivity is just very low.
The 24 hours continuous light curve coverage by OGLE, MOA and KMTNet \citep{kim16} will increase the number of detections of multiple cold planets systems.
The ultimate light curve coverage by {\it WFIRST} in terms of continuous coverage and high precision photometry 
is expected to dramatically increase the detections of multiple systems \citep{ben02}, as its sensitivity is much higher for both the mass ratio and separation parameter space. 
The precisely measured occurrence rate of multiple systems can be directly compared with the population synthesis simulations \citep[e.g.]{ida04,mor09}.
The statistical studies with the larger sample will allow us to put constrain on the planet formation models 
from the point of view of multiple cold planets.

\acknowledgments

D.S. and D.P.B. acknowledge support from NASA grants NNX13AF64G, NNX14AG49G,
and NNX15AJ76G.
T.S. acknowledges the financial support from JSPS23103002, JSPS24253004 and JSPS26247023. 
A.Y. acknowledges the financial support from JSPS25870893.
The MOA project is supported by grants JSPS25103508, JSPS23340064 and JP17H02871.
OGLE Team thanks Profs. M. Kubiak and G. Pietrzy{\'n}ski, for their
contribution to the collection of the OGLE photometric data.
The OGLE project has received funding from the National Science Centre,
Poland, grant MAESTRO 2014/14/A/ST9/00121 to A.U. 
Work  by  C.  Han  was  supported  by  the  grant 
(2017R1A4A1015178)  of  National  Research  Foundation of Korea.

\clearpage



\clearpage

\clearpage

\clearpage

\clearpage

\clearpage

\clearpage

\clearpage

\clearpage

\begin{figure}
\begin{center}
\plotone{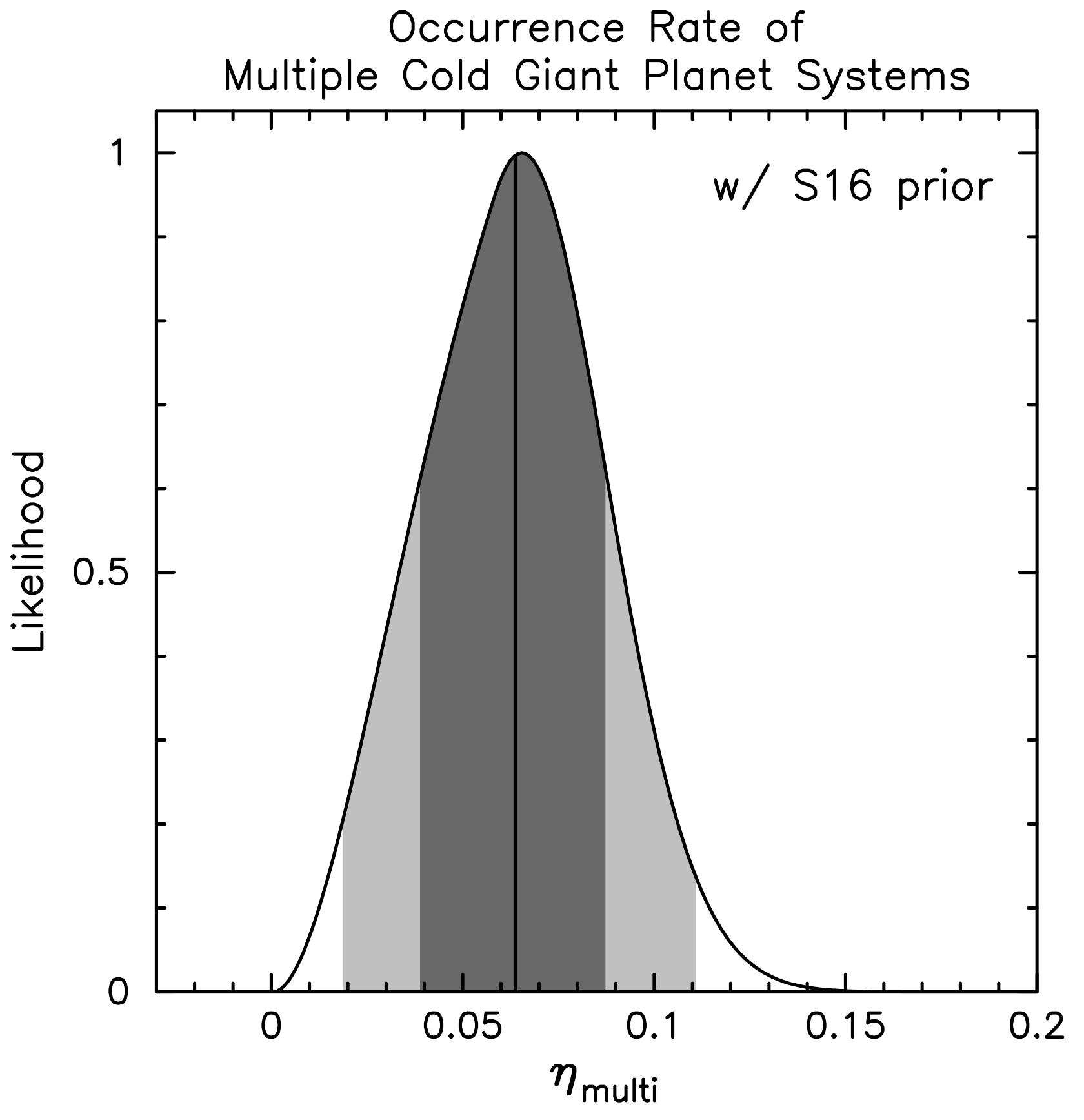}
\caption{Same as Figure \ref{fig:occ_mode}, but the average number of cold gas giants from \citet{suz16} is used for an upper limit for a prior distribution of $\eta_{\rm multi}$.
}
\label{fig:occ_mode_prior}
\end{center}
\end{figure}

\clearpage


\begin{table}
\caption{Best Fit Parameters}
\label{tab:bestfit}
\begin{center}
\small{

\begin{tabular}{lcccccc}
\tableline \tableline
Parameters & Best Close & MCMC Close & Best Wide & MCMC Wide & Best 1-Planet & MCMC 1-Planet\\
\tableline

$\chi^{2} $		& 4191.0		& \nodata  & 4192.2	& \nodata	& 4361.7  & \nodata\\
$t_{0}\ (\rm HJD')$		& 6900.224 & $6900.228_{-0.008}^{+0.009}$	& 6900.215 & $6900.219_{-0.011}^{+0.012}$  & 6900.191 & $6900.191_{-0.010}^{+0.010}$\\
$t_{\rm E}\ (\rm days)$ 	& 23.819 & $23.797_{-0.380}^{+0.418}$ 	& 23.867 & $23.909_{-0.582}^{+0.801}$ 	& 24.211 & $24.234_{-0.512}^{+0.513}$  \\ 
$u_{0}$ 				& -0.131 & $-0.132_{-0.003}^{+0.003}$	& -0.131 & $-0.131_{-0.003}^{+0.004}$	& 0.127 & $0.127_{-0.003}^{+0.003}$ \\
$q\ (10^{-4})$			& 4.468 & $4.451_{-0.557}^{+0.693}$ 	& 4.146 & $4.263_{-0.607}^{+0.678}$	&4.335 & $4.534_{-0.583}^{+0.705}$  \\
$s$ 					& 0.754 & $0.754_{-0.004}^{+0.004}$	& 0.754 & $0.754_{-0.005}^{+0.006}$	& 0.757 & $0.757_{-0.004}^{+0.004}$ \\
$\alpha\ (\rm rad)$ 		& 0.228 & $0.230_{-0.024}^{+0.020}$	& 0.239 & $0.242_{-0.032}^{+0.035}$ 	& 0.209 & $0.205_{-0.025}^{+0.027}$ \\
$\rho$ 				& 0.005 & $0.007_{-0.005}^{+0.006}$	& 0.002 & $0.006_{-0.004}^{+0.005}$	& 0.007 & $0.008_{-0.005}^{+0.006}$ \\
$\pi_{\rm E, N}$		& 0.199 & $0.189_{-0.553}^{+0.570}$  	& -0.077 & $-0.153_{-0.924}^{+0.843}$	& -0.585 & $-0.665_{-0.595}^{+0.655}$	\\ 
$\pi_{\rm E, E}$			& 0.092 & $0.094_{-0.083}^{+0.101}$	& 0.103 & $0.104_{-0.101}^{+0.110}$	& 0.002 & $-0.010_{-0.083}^{+0.086}$ 	\\
$q_{2}\ (10^{-4})$	 	& 6.388 & $7.071_{-1.832}^{+2.371}$	& 6.245 & $7.098_{-1.855}^{+1.836}$ 	& \nodata & \nodata \\
$s_{2}$		 		& 0.851 & $0.843_{-0.030}^{+0.034}$	& 1.358 & $1.368_{-0.044}^{+0.044}$ 	& \nodata & \nodata \\
$\psi\ (\rm rad)$ 		& 2.196 & $2.190_{-0.033}^{+0.028}$ 	& 2.207 & $2.209_{-0.036}^{+0.036}$ 	& \nodata & \nodata \\
\tableline
$I_{\rm S}$ (mag)		& 18.88	 &	$18.87 \pm 0.03$		& 18.87	& $18.88 \pm 0.04$			& 18.92 	& $18.92 \pm 0.04$	\\
$f_{\rm S}$			&  1.16	 &	$1.16 \pm 0.03$		& 1.16	& $1.16 \pm 0.04 $			& 1.11	& $1.11 \pm 0.04$\\
\tableline
\end{tabular}

\tablecomments{
$\rm HJD' \equiv$  HJD - 2450000. $f_{\rm S} = F_{\rm S}/(F_{\rm S} + F_{\rm B})$ is the blending parameter. 
The Best Close values are plotted as plus marks in Figure \ref{fig:chain}. 
}

}
\end{center}
\end{table}
\clearpage

\begin{table}
\caption{Parameters for the Binary Source Model}
\label{table-BS}
\begin{center}

\begin{tabular}{lc}
\tableline \tableline
Parameters & Values \\
\tableline


$\chi^{2}$  &  $4196.7$   \\
$t_{0}\ (\rm HJD')$   		& $6900.244_{-0.001}^{+0.001}$  \\
$t_{\rm E}\, (\rm days)$ 	& $23.893_{-0.227}^{+0.202}$    \\
$u_{0}$                  		& $0.133 \pm 0.002$   \\
$q\ (10^{-4})$         		& $4.218_{-0.267}^{+0.285}$  \\
$s$                        		& $0.753 \pm 0.002$  \\
$\alpha\ (\rm rad)$    		& $ 0.238 \pm 0.002 $  \\
$\rho \ (10^{-3})$            	& $0.709_{-0.272}^{+0.555}$ \\
$t_{0,2}\ (\rm HJD')$      	& $ 6898.960_{-0.003}^{+0.006} $ \\
$u_{0,2}\,(10^{-3})$       	& $ 7.000_{-0.235}^{+0.216}$ \\
$f_{\rm ratio}\,(10^{-3})$  	& $3.575_{-1.160}^{+0.320}$ \\
\tableline
$I_{\rm S1}\,(\rm mag)$	& $ 18.88 \pm 0.01$ \\
$I_{\rm S2}\,(\rm mag)$	& $ 25.00 \pm 0.23$ \\
$f_{\rm S12} $			& $ 1.16 \pm 0.02$ \\
\tableline
\end{tabular}

\tablecomments{
$\rm HJD' \equiv$  HJD - 2450000. 
$I_{\rm S1}$ and $I_{\rm S2}$ are apparent magnitudes of the primary and companion source stars.
$f_{\rm S12} = F_{\rm S12}/(F_{\rm S12} + F_{\rm B})$, where $F_{\rm S12}$ is the total flux from the source system, is the blending parameter. 
}

\end{center}
\end{table}
\clearpage

\begin{table}
\caption{Physical Parameters for Lens System}
\label{tab:physical}
\begin{center}

\begin{tabular}{lccc}
\tableline \tableline
Parameters & Units & \multicolumn{2}{c}{Values}\\
 & & Close & Wide \\
\tableline


$D_{\rm L}$         & kpc              &  $6.40_{-1.87}^{+1.28}$  & $6.38_{-1.93}^{+1.29}$ \\
$M_{\rm host}$    & $M_\odot$   & $0.40_{-0.24}^{+0.36}$  &  $0.39_{-0.24}^{+0.36}$\\
$m_{\rm b}$         & $M_\oplus$ & $55.3_{-33.1}^{+51.3}$  & $55.0_{-33.2}^{+51.4}$  \\
$m_{\rm c}$         & $M_\oplus$ & $83.7_{-51.2}^{+86.4}$  & $83.3_{-51.3}^{+86.6}$ \\
$r_{\perp, \rm b}$ & AU              & $1.5 \pm 0.6$ &  $1.5 \pm 0.6$ \\
$r_{\perp, \rm c}$ & AU              & $1.7_{-0.6}^{+0.7}$ & $2.7_{-1.0}^{+1.1}$ \\
$a_{\rm 3D, b}$         & AU              & $1.9_{-0.8}^{+1.3}$ & $1.9_{-0.8}^{+1.3}$ \\
$a_{\rm 3D, c}$         & AU              & $2.5_{-1.1}^{+2.2}$ & $3.5_{-1.4}^{+2.4}$ \\
\tableline
\end{tabular}


\end{center}
\end{table}
\clearpage

\clearpage


\clearpage



\begin{thebibliography}{}

\bibitem[Albrow et al.(2000)]{alb00} Albrow, M.~D., Beaulieu, J.-P., Caldwell, J.~A.~R., et al.\ 2000, \apj, 534, 894 
\bibitem[Bachelet et al.(2012)]{bac12} Bachelet, E., Shin, I.-G., Han, C., et al.\ 2012, \apj, 754, 73
\bibitem[Batista et al.(2015)]{bat15} Batista, V., Beaulieu, J.-P., Bennett, D.~P., et al.\ 2015, \apj, 808, 170
\bibitem[Batista et al.(2014)]{bat14} Batista, V., Beaulieu, J.-P., Gould, A., et al.\ 2014, \apj, 780, 54
\bibitem[Beaulieu et al.(2016)]{jp16} Beaulieu, J.-P., Bennett, D.~P., Batista, V., et al.\ 2016, \apj, 824, 83
\bibitem[Beaulieu et al.(2006)]{jp06} Beaulieu, J.-P., Bennett, D.~P., Fouqu{\'e}, P., et al.\ 2006, \nat, 439, 437r
\bibitem[Bennett(2010)]{bennett-himag} Bennett, D.P.\ 2010, \apj, 716, 1408
\bibitem[Bennett \& Rhie(1996)]{ben96} Bennett, D.~P., \& Rhie, S.~H.\ 1996, \apj, 472, 660 
\bibitem[Bennett \& Rhie(2002)]{ben02} Bennett, D.~P., \& Rhie, S.~H.\ 2002, \apj, 574, 985
\bibitem[Bennett et al.(1999)]{ben99} Bennett, D.~P., Rhie, S.~H., Becker, A.~C., et al.\ 1999, \nat, 402, 57 
\bibitem[Bennett et al.(2010)]{ben10} Bennett, D.~P., Rhie, S.~H., Nikolaev, S., et al.\ 2010, \apj, 713, 837 
\bibitem[Bennett et al.(2006)]{ben06} Bennett, D.~P., Anderson, J., Bond, I.~A., Udalski, A., \& Gould, A.\ 2006, ApJL, 647, L171
\bibitem[Bennett et al.(2007)]{ben07} Bennett, D.~P., Anderson, J., \& Gaudi, B.~S.\ 2007, \apj, 660, 781
\bibitem[Bennett et al.(2015)]{ben15hst} Bennett, D.~P., Bhattacharya, A., Anderson, J., et al., 2015, ApJ, 808, 169
\bibitem[Bennett et al.(2016)]{ben16} Bennett, D.~P., Rhie, S.~H., Udalski, A., et al.\ 2016, \aj, 152, 125
\bibitem[Bennett et al.(2012)]{ben12} Bennett, D.~P., Sumi, T., Bond, I.~A., et al.\ 2012, \apj, 757, 119 
\bibitem[Bennett et al.(2017)]{ben17} Bennett, D.~P., et al. 2017 in prep.
\bibitem[Bensby et al.(2011)]{bens11} Bensby, T., Ad\'{e}n, D., Mel\'{e}ndez, J., et al. 2011, A\&A, 533, A134
\bibitem[Bhattacharya et al.(2017)]{apa17} Bhattacharya, A., Bennett, D.~P., Anderson, J., et al.\ 2017, \aj, 154, 59
\bibitem[Bond et al.(2001)]{bon01} Bond, I. A., Abe, F., Dodd, R. J., et al. 2001, \mnras, 327, 868
\bibitem[Bond et al. (2017)]{bon17} Bond, I. A., et al. 2017, arXiv: 1703.08639
\bibitem[Bramich(2008)]{bra08} Bramich, D.~M.\ 2008, \mnras, 386, L77
\bibitem[Bressan et al.(2012)]{bre12} Bressan, A., Marigo, P., Girardi, L., et al.\ 2012, \mnras, 427, 127 
\bibitem[Cardelli et al.(1989)]{car89} Cardelli, J.~A., Clayton, G.~C., \& Mathis, J.~S.\ 1989, \apj, 345, 245
\bibitem[Cassan et al.(2012)]{cas12} Cassan, A., Kubas, D., Beaulieu, J.-P., et al.\ 2012, Nature, 481, 167
\bibitem[Chen et al.(2015)]{che15} Chen, Y., Bressan, A., Girardi, L., et al.\ 2015, \mnras, 452, 1068 
\bibitem[Chen et al.(2014)]{che14} Chen, Y., Girardi, L., Bressan, A., et al.\ 2014, \mnras, 444, 2525 
\bibitem[Chung et al.(2005)]{chu05} Chung, S.-J., Han, C., Park, B.-G., et al.\ 2005, \apj, 630, 535
\bibitem[Fabrycky et al.(2014)]{fab14} Fabrycky, D.~C., Lissauer, J.~J., Ragozzine, D., et al.\ 2014, \apj, 790, 146
\bibitem[Fukui et al.(2015)]{fuk15} Fukui, A., Gould, A., Sumi, T., et al.\ 2015, \apj, 809, 74 
\bibitem[Gaudi(2012)]{gau12} Gaudi, B.~S.\ 2012, ARA\&A, 50, 411
\bibitem[Gaudi(1998)]{gau98_BS} Gaudi, B.~S.\ 1998, \apj, 506, 533 
\bibitem[Gaudi et al.(1998)]{gau98_multi} Gaudi, B.~S., Naber, R.~M., \& Sackett, P.~D.\ 1998, \apjl, 502, L33
\bibitem[Gaudi et al.(2008)]{gau08} Gaudi, B.~S., Bennett, D.~P., Udalski, A., et al.\ 2008, Science, 319, 927
\bibitem[Gillon et al.(2017)]{gil17} Gillon, M., Triaud, A.~H.~M.~J., Demory, B.-O., et al.\ 2017, \nat, 542, 456 
\bibitem[Gladman(1993)]{gla93} Gladman, B.\ 1993, \icarus, 106, 247 
\bibitem[Gonzalez et al.(2012)]{gon12} Gonzalez, O.~A., Rejkuba, M., Zoccali, M., et al.\ 2012, \aap, 543, A13
\bibitem[Gould(2000)]{gou00} Gould, A.\ 2000, \apj, 542, 785
\bibitem[Gould \& Loeb(1992)]{gou92} Gould, A., \& Loeb, A.\ 1992, \apj, 396, 104
\bibitem[Gould et al.(2010)]{gou10} Gould, A., Dong, S., Gaudi, B. S., et al. 2010, \apj, 720, 1073
\bibitem[Griest \& Safizadeh(1998)]{gri98} Griest, K., \& Safizadeh, N.\ 1998, \apj, 500, 37 
\bibitem[Han(2005)]{han05} Han, C.\ 2005, \apj, 629, 1102
\bibitem[Han(2006)]{han06} Han, C.\ 2006, \apj, 638, 1080 
\bibitem[Han et al.(2016)]{han16} Han, C., Bennett, D.~P., Udalski, A., \& Jung, Y.~K.\ 2016, \apj, 825, 8
\bibitem[Han et al.(2001)]{han01} Han, C., Chang, H.-Y., An, J.~H., \& Chang, K.\ 2001, \mnras, 328, 986 
\bibitem[Han \& Park(2002)]{hanpark02} Han, C., \& Park, M.-G.\ 2002, Journal of Korean Astronomical Society, 35, 35
\bibitem[Han et al.(2013)]{han13} Han, C., Udalski, A., Choi, J.-Y., et al.\ 2013, \apjl, 762, L28
\bibitem[Han \& Gould(1995)]{han95} Han, C., \& Gould, A.\ 1995, \apj, 447, 53
\bibitem[Ida \& Lin(2004)]{ida04} Ida, S., \& Lin, D.~N.~C.\ 2004a, ApJ, 604, 388
\bibitem[Ida \& Lin(2005)]{ida05} Ida, S., \& Lin, D.~N.~C.\ 2005, \apj, 626, 1045  
\bibitem[Jung et al.(2017b)]{jun17blbs} Jung, Y.~K., Udalski, A., Bond, I.~A., et al.\ 2017b, \apj, 841, 75
\bibitem[Jung et al.(2017a)]{jun17bs} Jung, Y.~K., Udalski, A., Yee, J.~C., et al.\ 2017a, \aj, 153, 129
\bibitem[Kennedy \& Kenyon(2008)]{kk08} Kennedy, G. M., \& Kenyon, S. J. 2008, ApJ, 673, 502
\bibitem[Kim et al.(2016)]{kim16} Kim, S.-L., Lee, C.-U., Park, B.-G., et al.\ 2016, Journal of Korean Astronomical Society, 49, 37 
\bibitem[Koshimoto et al.(2014)]{kos14} Koshimoto, N., Udalski, A., Sumi, T., et al.\ 2014, \apj, 788, 128
\bibitem[Koshimoto et al.(2017a)]{kos17} Koshimoto, N., Udalski, A., Beaulieu, J.~P., et al.\ 2017a, \aj, 153, 1 
\bibitem[Koshimoto et al.(2017b)]{kos17_contami} Koshimoto, N., Shvartzvald, Y., Bennett, D.~P., et al.\ 2017b, \aj, 154, 3
\bibitem[Lissauer et al.(2011b)]{kep11} Lissauer, J.~J., Fabrycky, D.~C., Ford, E.~B., et al.\ 2011, \nat, 470, 53
\bibitem[Lissauer et al.(2011a)]{lis11} Lissauer, J.~J., Ragozzine, D., Fabrycky, D.~C., et al.\ 2011, \apjs, 197, 8
\bibitem[Marois et al.(2008)]{mar08} Marois, C., Macintosh, B., Barman, T., et al.\ 2008, Science, 322, 1348
\bibitem[Miyake et al.(2011)]{miy11} Miyake, N., Sumi, T., Dong, S., et al.\ 2011, \apj, 728, 120
\bibitem[Mordasini et al.(2009)]{mor09} Mordasini, C., Alibert, Y., \& Benz, W.\ 2009, \aap, 501, 1139
\bibitem[Morton \& Winn(2014)]{mor14} Morton, T.~D., \& Winn, J.~N.\ 2014, \apj, 796, 47 
\bibitem[Mr{\'o}z et al.(2017a)]{mro17} Mr{\'o}z, P., Udalski, A., Bond, I.~A., et al.\ 2017a, \aj, 154, 205
\bibitem[Mr{\'o}z et al.(2017b)]{mro17ffp} Mr{\'o}z, P., Udalski, A., Skowron, J., et al.\ 2017b, \nat, 548, 183
\bibitem[Muraki et al.(2011)]{mur11} Muraki, Y., Han, C., Bennett, D. P., et al. 2011, \apj, 741, 22
\bibitem[Nataf et al.(2013)]{nat13} Nataf, D.~M., Gould, A., Fouqu{\'e}, P., et al.\ 2013, \apj, 769, 88
\bibitem[Nishiyama et al.(2009)]{nis09} Nishiyama, S., Tamura, M., Hatano, H., et al.\ 2009, \apj, 696, 1407
\bibitem[Park et al.(2004)]{par04} Park, B.-G., DePoy, D.~L., Gaudi, B.~S., et al.\ 2004, \apj, 609, 166 
\bibitem[Penny et al.(2013)]{pen13} Penny, M.~T., Kerins, E., Rattenbury, N., et al.\ 2013, \mnras, 434, 2
\bibitem[Poleski et al.(2014)]{pol14} Poleski, R., Skowron, J., Udalski, A., et al.\ 2014, \apj, 795, 42 
\bibitem[Raghavan et al.(2010)]{rag10} Raghavan, D., McAlister, H.~A., Henry, T.~J., et al.\ 2010, \apjs, 190, 1 
\bibitem[Rattenbury et al.(2015)]{rat15}  Rattenbury, N.~J., Bennett, D.~P., Sumi, T., et al.\ 2015, \mnras, 454, 946 
\bibitem[Rattenbury et al.(2002)]{rat02} Rattenbury, N.~J., Bond, I.~A., Skuljan, J., \& Yock, P.~C.~M.\ 2002, \mnras, 335, 159
\bibitem[Rhie et al.(2000)]{rhi00} Rhie, S.~H., Bennett, D.~P., Becker, A.~C., et al.\ 2000, ApJ, 533, 378
\bibitem[Sako et al.(2008)]{sak08} Sako, T., Sekiguchi, T., Sasaki, M., et al. 2008, Experimental Astronomy, 22, 51
\bibitem[Schechter et al.(1993)]{dophot} Schechter, P.~L., Mateo, M., \& Saha, A.\ 1993, \pasp, 105, 1342
\bibitem[Shvartzvald et al.(2014)]{shv14} Shvartzvald, Y., Maoz, D., Kaspi, S., et al.\ 2014, \mnras, 439, 604
\bibitem[Shvartzvald et al.(2016)]{shv16} Shvartzvald, Y., Maoz, D., Udalski, A., et al.\ 2016, \mnras, 457, 4089 
\bibitem[Skowron et al.(2016)]{sko16} Skowron, J., Udalski, A., Poleski, R., et al.\ 2016, \apj, 820, 4 
\bibitem[Smith et al.(2007)]{smi07} Smith, M.~C., Wo{\'z}niak, P., Mao, S., \& Sumi, T.\ 2007, \mnras, 380, 805
\bibitem[Spergel et al.(2015)]{spe15} Spergel, D., Gehrels, N., Baltay, C., et al.\ 2015, arXiv:1503.03757 
\bibitem[Street et al.(2016)]{str16} Street, R.~A., Udalski, A., Calchi Novati, S., et al.\ 2016, \apj, 819, 93
\bibitem[Sumi et al.(2003)]{sum03}  Sumi, T., Abe, F., Bond, I. A., et al. 2003, \apj, 591, 204
\bibitem[Sumi et al.(2011)]{sum11} Sumi, T., Kamiya, K., Bennett, D. P., et al. 2011, Natur, 473, 349
\bibitem[Sumi et al.(2006)]{sum06} Sumi, T., Wo{\'z}niak, P.~R., Udalski, A., et al.\ 2006, \apj, 636, 240
\bibitem[Suzuki et al.(2014)]{suz14} Suzuki, D., Udalski, A., Sumi, T., et al.\ 2014, \apj, 780, 123
\bibitem[Suzuki et al.(2016)]{suz16} Suzuki, D., Bennett, D.~P., Sumi, T., et al.\ 2016, \apj, 833, 145
\bibitem[Street et al.(2016)]{str16} Street, R.~A., Udalski, A., Calchi Novati, S., et al.\ 2016, \apj, 819, 93 
\bibitem[Tang et al.(2014)]{tan14} Tang, J., Bressan, A., Rosenfield, P., et al.\ 2014, \mnras, 445, 4287
\bibitem[Udalski (2003)]{uda03} Udalski, A. 2003, \actaa, 53, 291
\bibitem[Udalski et al.(2015a)]{uda15} Udalski, A., Jung, Y.~K., Han, C., et al.\ 2015a, \apj, 812, 47
\bibitem[Udalski et al.(2015b)]{uda15ogle4} Udalski, A., Szyma{\'n}ski, M.~K., \& Szyma{\'n}ski, G.\ 2015b, \actaa, 65, 1
\bibitem[Udalski et al.(2015c)]{uda15_sp} Udalski, A., Yee, J.~C., Gould, A., et al.\ 2015c, \apj, 799, 237 
\bibitem[Verde et al.(2003)]{ver03} Verde, L., Peiris, H. V., Spergel, D. N., 2003, \apjs, 148, 195
\bibitem[Yee et al.(2012)]{yee12} Yee, J.~C., Shvartzvald, Y., Gal-Yam, A., et al.\ 2012, ApJ, 755, 102

\end{thebibliography}
\end{document}